\documentclass[aps,prd,longbibliography,nofootinbib,amsthm,amsmath,amssymb,amsfonts,notitlepage]{revtex4-1}
\usepackage[utf8]{inputenc}
\usepackage[T1]{fontenc}
\usepackage[english]{babel}
\usepackage{graphicx}
\usepackage{float}
\usepackage{xcolor}

\newcommand{\bk}{{\bf k}}
\newcommand{\bb}{{\bf b}}

\newcommand{\bP}{{\bf P}}
\newcommand{\bp}{{\bf p}}
\newcommand{\br}{{\bf r}}

\newcommand{\bbe}{{\bf e}}
\newcommand{\lr}[1]{ \langle #1 \rangle}

\def\lsim{\mathrel{\rlap{\lower4pt\hbox{\hskip1pt$\sim$}}
		\raise1pt\hbox{$<$}}}         

\def\gsim{\mathrel{\rlap{\lower4pt\hbox{\hskip1pt$\sim$}}
		\raise1pt\hbox{$>$}}}         

\date{\today}

\begin{document}

\title{Threshold effects in high-energy vortex state collisions}
\author{Bei Liu}
\email{liub98@mail2.sysu.edu.cn}
\author{Igor P. Ivanov}
\email{ivanov@mail.sysu.edu.cn}
\affiliation{School of Physics and Astronomy, Sun Yat-sen University, 519082 Zhuhai, China}

\begin{abstract}
Collisions of particles prepared in non--plane-wave states with a non-trivial phase structure, 
such as vortex states carrying an adjustable orbital angular momentum (OAM), 
open novel opportunities in atomic, nuclear, and 
high-energy physics unavailable for traditional scattering experiments.
Recently, it was argued that photoinduced processes such as
$\gamma d \to pn$ and $\gamma p \to \Delta^+$ initiated by a high-energy vortex photon 
should display a remarkable threshold shift and a sizable cross section enhancement
as the impact parameter $b$ of the target hadron with respect to the vortex photon axis goes to zero.
In this work, we theoretically explore whether this effect exists within the quantum-field-theoretic
treatment of the scattering process. We do not rely on the semiclassical assumption
of pointlike, non-spreading target particle and, instead, consider the toy process of heavy particle pair production 
in collision of two light particles prepared as a Laguerre-Gaussian and a compact Gaussian wave packets,
paying special attention to the threshold behavior of the cross section.
We do observe threshold smearing due to non-monochromaticity of the wave packets,
but we do not confirm the near-threshold enhancement. 
Instead we find an OAM-related dip at $b\to 0$ as compared with 
the two Gaussian wave packet collision.
\end{abstract}

\maketitle

\section{Introduction}

\subsection{Vortex states and their collisions}

When describing a high-energy collision process, one usually represents the initial state particles as plane waves.
Although real collisions take place between localized wave packets,
the effects of localization are unimportant in most cases, 
and the limit of an infinitely narrow momentum space wave function can be taken, see e.g. \cite{Peskin:1995ev}.
Very few examples exist \cite{Kotkin:1992bj} of collider processes in which the non--plane-wave nature 
of the initial particles plays a role.

Recently, a new direction of research emerged, in which one studies collisions of photons, electrons, 
and other particles prepared in the so-called vortex states, see \cite{Ivanov:2022jzh} for a recent review
(extension to other structured wave packets with non-trivial phase configurations can be found in \cite{Karlovets:2016jrd}).
A vortex state is described by the coordinate space wave function which, 
apart from the usual phase factor $\exp(-i Et + i k_z z)$, 
depends on the azimuthal angle $\varphi$ via $\exp(i\ell \varphi)$, with an integer winding number $\ell$.
Put simply, not only does this state propagate along axis $z$ but it also rotates around its average propagation direction.
A one-particle vortex state possesses a non-zero intrinsic orbital angular momentum (OAM) projection, 
which can be adjusted at will.
Such states have been experimentally demonstrated for photons \cite{Allen:1992zz,Bahrdt:2013eoa,photonics-review-2017,Nature-Phot-2019}, 
electrons \cite{Bliokh:2007ec,Uchida:2010,Verbeeck:2010,McMorran:2011}, 
cold neutrons \cite{clark2015controlling,sarenac2019generation,Sarenac:2022}, and slow atoms \cite{luski2021vortex}.
Although the energy range available so far is limited, there exist numerous suggestions of bringing vortex states
into the high-energy domain. More information on the present experimental situation and future prospects
can be found in reviews \cite{Ivanov:2022jzh,Paggett:2017,Knyazev-Serbo:2018,Bliokh:2017uvr,Lloyd:2017,larocque2018twisted,sarenac2018methods}. 

In particle and nuclear physics, the intrinsic OAM of the initial state represents a new, previously unexplored degree of freedom, 
and it is interesting to understand what insights collisions of such states can offer \cite{Ivanov:2022jzh}.
Theoretical investigation of high-energy collisions of vortex states started more than a decade ago
\cite{Jentschura:2010ap,Jentschura:2011ih,Ivanov:2011kk,Karlovets:2012eu}. 
It has already led to predictions of remarkable effects
such as novel features of the Schwinger scattering of slow neutrons on nuclei \cite{Afanasev:2019rlo,Afanasev:2021uth},
sensitivity to the overall phase of the scattering amplitude \cite{Ivanov:2012na,Ivanov:2016oue,Karlovets:2016dva,Karlovets:2016jrd}, 
and novel spin effects induced by the intrinsic spin-orbital interactions within vortex beams \cite{Ivanov:2019vxe,Ivanov:2020kcy}.
Additional insights into the proton structure are to be expected in deep inelastic scattering with vortex electrons and protons.

Recently, another remarkable effect was proposed in \cite{Afanasev:2020nur,Afanasev:2021fda}.
Building on previous atomic physics \cite{barnett2013superkick} and nuclear physics \cite{Afanasev:2017jdf,Afanasev:2019rlo} studies, 
the authors considered absorption of high-energy vortex photon by a pointlike target 
placed at distance $b$ from the phase singularity line. 
In this regime, one expects the so-called superkick effect \cite{barnett2013superkick,Afanasev:2020nur},
a surprisingly large transverse momentum transfer to be explained in detail below.
The authors of \cite{Afanasev:2017jdf,Afanasev:2019rlo} predicted that, 
apart from mere modification of the transverse momentum distribution, this effect 
should also significantly change the total cross section in the vicinity of the threshold.
Calculations of the deuteron vortex photodisintegration process $\gamma d\to pn$ \cite{Afanasev:2020nur}
and the vortex photoproduction of $\Delta$ resonance \cite{Afanasev:2021fda}
showed a shift of the threshold and a dramatic enhancement of the cross section as $b \to 0$.
These total cross section modifications, if real, would constitute a spectacular experimental signal.

\subsection{The superkick phenomenon and its analyses}

\begin{figure}[!h]
	\centering
	\includegraphics[width=0.3\textwidth]{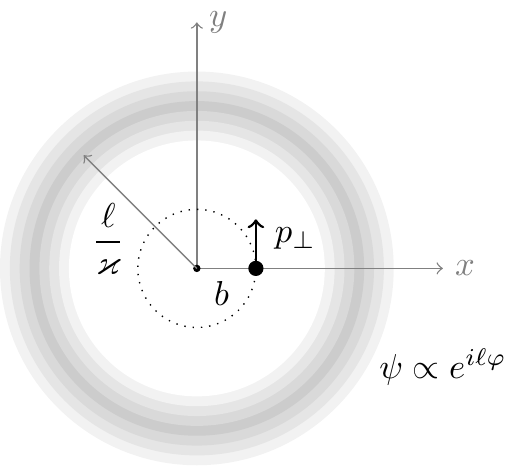}
	\caption{An atom placed at distance $b \ll \ell/\varkappa$ from the Bessel photon axis can probes 
	the local transverse momentum $p_\perp \gg \varkappa$.}
	\label{fig-semiclassical}
\end{figure}

To get an intuitive understanding of the superkick effect,
consider a pointlike probe (an atom) in the light field of an optical vortex,
which, for the moment, we describe as a Bessel vortex state. 
In the paraxial limit, one can approximate the four-potential of the optical vortex in the transverse plane 
as a constant polarization vector $e^\mu$ times a Bessel-state wave function: 
\begin{equation}
A^\mu_{\varkappa, \ell}(\br_\perp) \propto e^\mu e^{i\ell\varphi} J_{|\ell|}(\varkappa r_\perp)\,.
\end{equation}
Here, the axis $r_\perp = 0$ corresponds to the phase singularity: 
the phase is undefined and, therefore, the intensity must vanish.
As a result, the transverse intensity profile has the shape of concentric rings, 
with the first one of radius $\ell/\varkappa$, see Fig.~\ref{fig-semiclassical}.
Interpreted in quantum terms, this Bessel light field is represented by a coherent superposition of
plane wave photons with the transverse momenta $\bk_\perp$ of equal absolute values $\varkappa$ coming from all azimuthal angles. 

A pointlike probe placed at distance $b$ from the phase singularity line 
experiences	 the phase gradient $\ell/b$, which can be interpreted as the local momentum \cite{berry2013five},
see also the recent discussion \cite{Afanasev:2022vgl}.
If we choose the unit vectors $\bbe_x$ and $\bbe_y$ on the transverse plane and consider the
point $\bb = b\, \bbe_x$, then the local transverse phase gradient produces the local momentum orthogonal to $\bb$:
\begin{equation}
\bp_\perp = \frac{\ell}{b}\, \bbe_y\label{mb}\,.
\end{equation}
For a sufficiently small $b$, this local momentum can be arbitrarily large, 
much larger than the photon transverse momentum $\varkappa$. 
We arrive at a paradox: a photon absorbed by the probe exerts a much larger momentum transfer 
than it actually carries.

This surprisingly large momentum transfer was dubbed \cite{barnett2013superkick} the ``superkick''.
As paradoxical as it may seem in the semiclassical picture with a pointlike probe,
it finds its natural resolution within the quantum treatment, 
which was first outlined in \cite{barnett2013superkick}.
When the probe atom is represented by a compact wave packet
of spatial extent $\sigma \ll b$, then its momentum space wave function must extend up 
to values of about $1/\sigma \gg 1/b$.
In the initial wave packet, all plane wave components balance each other, which leads to $\langle\bk_\perp\rangle = 0$.
Absorption of a photon disturbs the balance leading to a non-zero average transverse momentum, which can be large.
Thus, the role of the photon in this process is not to supply a large transverse momentum to the atom, 
but only to trigger a momentum bias among the plane wave components, and this bias results in the superkick.

The superkick effect may also lead to novel insights in nuclear and high-energy physics,
and remarkable additional effects were indeed proposed in 
\cite{Afanasev:2020nur,Afanasev:2021fda}.
In order to discuss the effect in particle collisions,
one first needs to rederive it within the full quantum field theoretical approach.
This was first done in our paper \cite{Ivanov:2022sco}, 
where the process was formulated as a collision 
of a vortex Bessel state possessing the conical momentum $\varkappa$
with a monochromatic Gaussian beam of the longitudinal momentum $p_z$ and 
a transverse localization $\sigma$, shifted by the impact parameter $b$ from the vortex axis.
For this multi-scale problem, we showed that the superkick phenomenon 
can indeed be observable if the following hierarchy of scales is respected: 
\begin{equation}
\frac{1}{p_z} \ll \sigma \ll b \ll \frac{\ell}{\varkappa}\,.\label{scales}
\end{equation}
In accordance with \cite{barnett2013superkick}, 
we demonstrated that the superkick phenomenon cannot be properly understood 
without taking into account the transverse localization and the evolution of the wave packets.
In the naive picture of a pointlike target particle, the superkick would be completely mysterious.

\subsection{Energy threshold modification due to the superkick effect?}

While the superkick effect in atomic physics still awaits experimental confirmation,
its existence and origin were verified beyond any doubt
within different approaches \cite{barnett2013superkick,Afanasev:2020nur,Afanasev:2021fda,Ivanov:2022sco}.
The next question is whether the same kinematical regime leads to additional observable effects,
especially in nuclear and hadronic collisions.

In the first study which addressed this question \cite{Afanasev:2020nur}, 
the deuteron photodisintegration process $\gamma d\to pn$ was considered
within the superkick kinematics, with the vortex gamma photon energy just above the photodisintegration threshold.
For the plane wave photon, the nucleons $p$ and $n$ emerging from deuteron disintegration would be very slow.
However a vortex gamma photon carrying the same energy would exert a superkick on a pointlike deuteron 
placed near the vortex singularity axis. 
The authors of \cite{Afanasev:2020nur} predicted that the superkick-induced recoil should provide
an extra kinetic energy, which would significantly shift the threshold behavior of the cross section
and modify the energies of the final nucleons.
A similar dramatic threshold modification was also predicted in \cite{Afanasev:2021fda}
for the photoproduction of the $\Delta$ resonance on the proton. 

Observation of a dramatic change in the energy threshold behavior of the cross section 
would constitute a spectacular hadronic effect and a novel probe into hadronic dynamics.
However, the predictions of \cite{Afanasev:2020nur,Afanasev:2021fda} are difficult to reconcile 
with another argument.
A vortex state, in essence, is a superposition of many plane waves. 
If the cross section induced by each plane wave component is small, 
it is hard to imagine how it could become large for its superposition, the vortex state.

Since \cite{Afanasev:2020nur,Afanasev:2021fda} treated the target hadron semiclassically,
as a pointlike object rather than a spreading wave packet,
a reanalysis of this process is required in the standard collider-like kinematics
with freely propagating, spreading wave packets of certain initial size.

\subsection{The goals of the present study}

The main goal of the present work is to compute, within the superkick kinematics, a typical $2\to 2$ scattering process
which possesses an energy threshold and to check its near-threshold behavior.
This work can be viewed as a follow-up paper of \cite{Ivanov:2022sco}
and also as a quantum-field-theoretical testbed of some of the predictions 
of \cite{Afanasev:2020nur,Afanasev:2021fda}.

It turns out, however, that the formalism used in our previous work \cite{Ivanov:2022sco}
is not suitable for the energy behavior computation.
In \cite{Ivanov:2022sco}, we considered the collision of a vortex state modeled by a monochromatic Bessel beam
and a compact probe state described as a tightly focused Gaussian beam.
However, under these assumptions, the two beams never decouple from each other along the longitudinal axis,
even when they spread far away from the focal planes. As a result, the superkick effect essentially vanishes.
However, disappearance of the superkick is not a real physical effect but is an artefact
of the unrealistic assumption of the exact Bessel beam, the difficulty known since first papers
on vortex-state scattering.
Thus, an additional longitudinal regularization procedure was employed in \cite{Ivanov:2022sco}
to remove this assumption and to mimic the realistic case of wave packets of finite size.
With this regularization procedure, the superkick effect was indeed recovered.
However, the artificial regularization prevented us from analyzing the energy threshold behavior of the cross section.
Thus, the calculations of \cite{Ivanov:2022sco} cannot be considered fully satisfactory 
and should be replaced with an analysis of localized wave packet scattering.

In the present paper, on the way to reaching our main goal, we adapt the formalism \cite{Ivanov:2022sco} 
to realistic wave packet collisions of finite transverse and longitudinal size.
There exist several ways to do it.
The approach which we use is to replace the Bessel vortex states with the Laguerre-Gaussian (LG) wave packets.
Building on the earlier work \cite{Kotkin:1992bj},
the recent papers \cite{Karlovets:2018iww,Karlovets:2020odl} developed the Wigner function formalism 
for non-relativistic and relativistic scattering of Gaussian and LG wave packets, 
both within the paraxial approximation and beyond.
Although this approach can deal with a broad variety of situations \cite{Karlovets:2020odl},
we find it less intuitive than the one involving wave functions,
and we believe it is not mandatory for discussion of the effect under interest.

To this end, we recast the LG-based formalism of \cite{Karlovets:2018iww,Karlovets:2020odl}
back in the wave function language, apply it to scattering in the superkick kinematical regime,
and, once again, reconfirm the existence and properties of the superkick effect 
without resorting to any artificial regularization procedure.
This is, of course, ot a new result; we just view it as a cross-check of the validity of
our new formalism.

Equipped with this method, we will be able to track the energy behavior 
of the toy process (production of two heavy particles in collisions of two light ones)
in the threshold region.
We will compare the plane-wave results with the two Gaussian wave packet
and with the LG vs. Gaussian wave packet collisions at various impact parameters,
which will help us address the above controversy.

We stress that, in this study, we focus on the matter-of-principle effects
and do not discuss their experimental verification. 
The results of \cite{Afanasev:2020nur,Afanasev:2021fda} make it clear 
that verification would be very challenging due to the extreme focusing and localization of the wave packets.
What we discuss here is which effects exist in principle and which do not.

The structure of this paper is the following. 
In the next Section, we build the formalism suitable for the LG vs. Gaussian wave packet collision
analysis in the superkick regime. 
We begin with general expressions, discuss the time evolution of the collision event,
and define the impulse approximation 
(the collision happens on a much shorter timescale than the wave packet spreading).
In Section~\ref{section-main-results}, we apply the formalism to our process.
We first show that, far above the threshold, we approach the plane-wave cross section for arbitrarily shaped
wave functions, at least in the paraxial limit. We also reproduce the superkick effect.
Then we study the energy behavior of the cross section near the threshold, analytically and numerically,
and find that the $b$-dependence of the cross section does not confirm the threshold enhancement 
predicted in \cite{Afanasev:2020nur,Afanasev:2021fda}.
In the final section we discuss the results and draw conclusions.
The Appendix contains technical details of the vortex amplitude calculation.

Throughout the paper, vectors are denoted with bold symbols,
transverse vectors carry the subscript $\perp$. When giving the absolute values of the vectors,
the bold symbols are dropped.
The relativistic units $\hbar = c = 1$ are used. 

\section{Collisions of a LG and a Gaussian wave packets} \label{section-nonplanewave}

\subsection{General expressions}

Since we aim to investigate kinematical distributions in wave packet collisions, 
we focus on the process $\phi\phi\to \Phi\Phi$, in which two light scalar particles $\phi$ of mass $m$ 
produce two heavy scalar particles $\Phi$ of mass $M > m$. In this way, we avoid spin-related complications
which are inessential to the present work.

Let us begin with the textbook case of the plane wave collisions.
The two initial particles have the three-momenta $\bk_1$, $\bk_2$ and energies $E_1$, $E_2$;
the two final particles are described with the three-momenta $\bk'_1$, $\bk'_2$ and energies $E'_1$, $E'_2$. 
The plane wave $S$-matrix element has the form
\begin{equation}
S_{PW}(k_1,k_2;k'_1,k'_2) = i(2\pi)^4\delta^{(4)}(k_1+k_2 - k_1'-k_2') {{\cal M} \over \sqrt{16 E_1 E_2 E'_1 E'_2}}\cdot N_{PW}^4\,.
\label{SPW}
\end{equation}
Here, ${\cal M}$ is the invariant amplitude calculated according to the Feynman rules 
and $N_{PW}$ is the plane wave normalization factor.
Normalizing the initial states as one particle per large volume $V$, we get $N_{PW} = 1/\sqrt{V}$.
The plane wave scattering cross section can then be written as
\begin{equation}
d\sigma_{PW} = \frac{(2\pi)^4\delta^{(4)}(k_1+k_2 - k_1'-k_2') |{\cal M}|^2}{4I_{PW}}\cdot \frac{d^3k_1'}{(2\pi)^3 2E_1'}\, 
\frac{d^3k_2'}{(2\pi)^3 2E_2'}\,.
\end{equation}
where $I_{PW} = \sqrt{(k_1 k_2)^2-m^4}$ is the flux invariant.
Clearly, if the initial $\bk_1$ and $\bk_2$ are known, the total final state momentum $\bk_1 + \bk_2 =  \bP = \bk_1' + \bk_2'$ is also fixed.
Performing the integrals over the final phase space in the center of motion frame, in which $E_1 = E_2 = E_0$, 
we get the well-known differential cross section:
\begin{equation}
d\sigma_{PW}(E_0) = \frac{|{\cal M}|^2}{256\pi^2 E_0}\frac{\beta\, d\Omega_1}{p_0}\,,
\quad \beta = \sqrt{1 - \frac{M^2}{E_0^2}}\,, \quad p_0 = \sqrt{E_0^2-m^2}\,.\label{sigma-tot-dOmega}
\end{equation}
Here, $\Omega_1$ refers to the solid angle of the final particle with momentum $\bk_1'$; 
the momentum of the second final particle is exactly the opposite.
If the invariant amplitude ${\cal M}$ does not depend on the angles,
the angular integration leads to 
\begin{equation}
\sigma_{PW}(E_0) = \frac{|{\cal M}|^2}{128\pi E_0}\frac{\beta}{\sqrt{E_0^2-m^2}}\,,\label{sigma-tot-PW}
\end{equation}
where we took into account the symmetry factor for the two identical particles in the final state.
The energy behavior of the total cross section displays the well-known threshold
at the $E_0 = M$ followed by the sharp growth proportional to $\beta$ and a high-energy decrease $\propto 1/E_0^2$.

Let us now assume that the two initial particles are prepared as localized wave packets.
Scattering theory of arbitrarily shaped, partially coherent beams
was developed in the paraxial approximation in \cite{Kotkin:1992bj}, 
extended recently beyond the paraxial approximation in \cite{Karlovets:2016jrd,Karlovets:2018iww,Karlovets:2020odl}.
A particular version of this general procedure was used previously 
to compute scattering of Bessel twisted particle 
\cite{Jentschura:2010ap,Jentschura:2011ih,Ivanov:2011kk,Karlovets:2012eu,Ivanov:2016oue,Karlovets:2016jrd,Karlovets:2020odl}.
In our previous work on this problem \cite{Ivanov:2022sco}, we also used monochromatic Bessel and Gaussian beams
of infinite longitudinal extent, which forced us to introduce an artificial regularization procedure.
In the present study, we remove the fixed energy assumption and consider the wave packets to be localized in all directions.

We describe the two initial particles 
as momentum space wave packets $\phi_1(\bk_1)$ and $\phi_2(\bk_2)$ normalized as
\begin{equation}
\int \frac{d^3 k}{(2\pi)^3}\, \frac{1}{2E(k)}\, |\phi_i(\bk)|^2 = 1\,.
\end{equation}
Notice the Lorentz-invariant normalization condition for the momentum wave functions, 
which we adopt following \cite{Karlovets:2020odl}.
The coordinate wave functions are defined by
\begin{equation}
\psi(\br,t) = \int \frac{d^3k}{(2\pi)^3\sqrt{2E(k)}}\,\phi(\bk)\, e^{i\bk\br - iE(\bk)t}\,,\quad
\int d^3r\, |\psi(\br,t)|^2 = 1\,.\label{psi-def}
\end{equation}
Since we now work with wave packets, the initial particles do not possess definite momenta or energies.
We can define the average momentum in each of the two colliding wave packets, $\bp_1 = \langle \bk_1\rangle$ and $\bp_2 = \langle \bk_2\rangle$. 
We consider the collision setting in which these average momenta are antiparallel to each other and define the common axis $z$,
with $p_{1z} > 0$, $p_{2z} < 0$. 
At this state, we do not require them to sum up to zero: $p_{1z} \not = |p_{2z}|$, although in a later section we will adopt this reference frame choice.

Throughout the paper, we work within the paraxial approximation: the typical transverse momenta of the two wave functions 
are assumed to be much smaller that $p_{1z}$ and $|p_{2z}|$. Going beyond the paraxial approximation 
is possible \cite{Karlovets:2016jrd,Karlovets:2018iww,Karlovets:2020odl}
but we believe it will not provide additional insights into the kinematical features of the superkick scattering regime.

The momentum space wave functions are constructed in a way similar to \cite{Karlovets:2020odl} with a few differences outlined below.
For the scalar vortex state with a definite OAM value\footnote{For negative values of $\ell$, 
the expressions are the same with $\ell$ replaced by $|\ell|$ everywhere apart from the $\exp(i\ell\varphi)$ factor.
They do not lead to any novel features, so we select $\ell > 0$ to simplify the notation.} $\ell > 0$ (particle 1), 
we use the relativistic LG principal\footnote{Going beyond the principal modes leads 
to additional complications which do not seem essential for the physics of the phenomena we consider.
However it may prove useful to analyze this case as well, which can be done with the aid of expressions
given in \cite{Karlovets:2020odl}.} mode:
\begin{equation}
\phi_1(\bk_1) = (4\pi)^{3/4}\sigma_{1\perp}\sqrt{\sigma_{1z}} \sqrt{2E_1}\, \frac{(\sigma_{1\perp}k_{1\perp})^{\ell}}{\sqrt{\ell!}}
\exp\left[ -\frac{k_{1\perp}^2 \sigma_{1\perp}^2}{2}-\frac{(k_{1z}-p_{1z})^2 \sigma_{1z}^2}{2} +i\ell\varphi_k\right] \,.
\end{equation}
Here, $\sigma_{1\perp}$ and $\sigma_{1z}$ are the transverse and longitudinal spatial extents of the coordinate space wave function.
If one requires the Gaussian factor to become spherically symmetric in the rest frame,
the longitudinal extent will be relativistically contracted in the moving frame,
and one has to assign $\sigma_{1z} = \sigma_{1\perp}/\bar\gamma_1$, where 
$\bar\gamma_1 = \varepsilon_1/m$, $\varepsilon_1 = \sqrt{p_1^2 + m^2}$.
We prefer not to limit ourselves to this choice; instead, we keep the two parameters $\sigma_{1\perp}$ and $\sigma_{1z}$ independent. 
This approach allows us to access different wave packet configurations
and, if needed, to smoothly interpolate between a LG beam with infinite $z$ extent and a compact wave packet.
In fact, in numerical calculations below we will use $\sigma_{iz} \gg \sigma_{i\perp}$ in order to 
expose the impact parameter dependent threshold effects.

\begin{figure}[!h]
	\centering
	\includegraphics[width=0.4\textwidth]{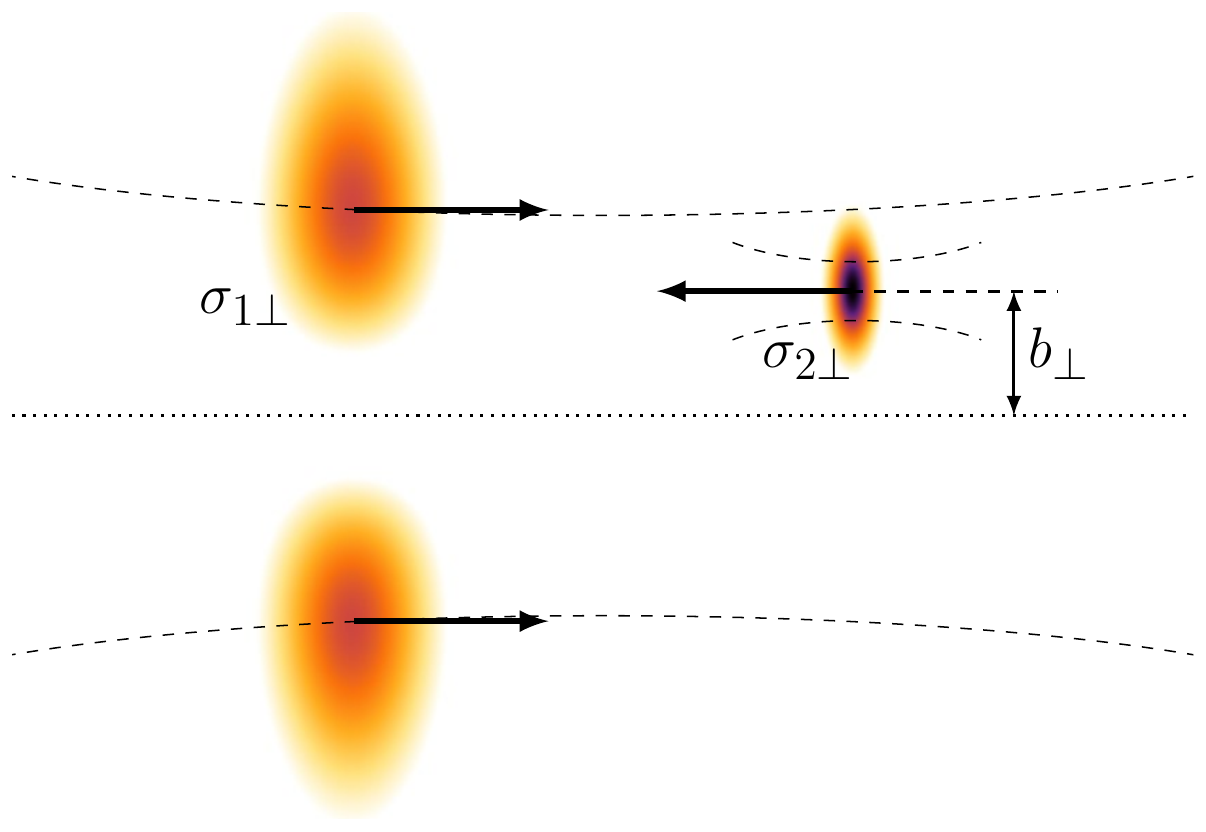}
	\caption{Collision of a wide LG wave packet shown by the two lobes on the left
	with a compact Gaussian wave packet on the right at a non-zero impact parameter $b \equiv |\bb_\perp|$.
	The dashed lines indicate the focusing and defocusing effect in wave packet evolution.}
	\label{fig-collision}
\end{figure}

The Gaussian state can be obtained from the above formula by setting $\ell = 0$. 
For particle 2, it is described by 
\begin{eqnarray}
\phi_2(\bk_2)&=&(4\pi)^{3/4}\sigma_{2\perp}\sqrt{\sigma_{2z}} \sqrt{2E_2}\,   
\exp\left[-\frac{k_{2\perp}^2 \sigma_{2\perp}^2}{2}-\frac{(k_{2z}-p_{2z})^2 \sigma_{2z}^2}{2} 
-i \bb_\perp \bk_{2\perp} -i b_z k_{2z} + i \tau E_2\right]\,,
\end{eqnarray}
with parameters $\sigma_{2\perp}$ and $\sigma_{2z}$.
Here, we take into account the possibility that the two wave packets may be shifted 
with respect to each other in three different ways, see Fig.~\ref{fig-collision}.
The impact parameter $\bb_\perp$ defines the transverse offset between their axes,
$b_z$ defines the longitudinal distance between their focal planes, 
and $\tau$ characterizes the time difference between the instants of their maximal focusing. 
In the following section we will demonstrate that $b_z$ and $\tau$ play a different role than $\bb_\perp$.

Following \cite{Kotkin:1992bj,Karlovets:2020odl}, we present the generalized cross section as
\begin{equation}
d\sigma = \frac{dW}{L}\,, \quad dW = (2\pi)^8 |{\cal I}|^2 \frac{d^3k_1'}{(2\pi)^3 2E_1'}\, \frac{d^3k_2'}{(2\pi)^3 2E_2'}\,.
\label{dsigma-WP}
\end{equation}
where
\begin{equation}
{\cal I} = \int\frac{d^3k_1}{(2\pi)^3 2E_1} \frac{d^3k_2}{(2\pi)^3 2E_2}\,\varphi_1(\bk_1)\,\varphi_2(\bk_2)
\delta^{(3)}(\bk_1+\bk_2 - \bP)\,\delta(E_1+E_2-E_f)\, {\cal M}\,.\label{cal-I}
\end{equation}
Since we describe the two final particles as plane waves, 
we have introduced their total momentum and total energy:
\begin{equation}
\bP = \bk'_1 + \bk'_2\,,\quad E_f = E_1'+E_2'\,.\label{totalPE}
\end{equation}
The quantity $L$ represents the Lorentz invariant luminosity for collision of two wave packets \cite{Kotkin:1992bj,Karlovets:2020odl}. 
Within the paraxial approximation, the relative velocity $|v_1-v_2|$ can be computed 
via the average momenta of the two wave packets $v_i = p_{iz}/\varepsilon_i$ (with $v_1>0$ and $v_2<0$)
and, consequently, taken out as a universal factor.
Then, the luminosity function represents the space-time overlap of the two colliding wave packets:
\begin{equation}
L = |v_1 - v_2| \int d^3 r\, dt\, |\psi_1(\br,t)|^2 |\psi_2(\br,t)|^2\,.\label{lumi}
\end{equation}
This factor takes care of the correct normalization which is especially important when two the colliding wave packets overlap only partially,
as is the case for a significant transverse offset $\bb_\perp$.

It is instructive to mention that in the semiclassical approximation in which one models the target particle
as a tightly confined, non-spreading wave packet located at position $\bb$, 
one can safely replace $\int d^3 r\, |\psi_1(\br,t)|^2 |\psi_2(\br,t)|^2 = |\psi_1(\bb,t)|^2$.
As a result, one obtains the (time-integrated) local flux of the first incident particle computed 
at the position of the target particle. Thus, in an appropriate limit,
our definition of flux (and, consequently, of the cross section)
matches the local flux definition used in \cite{Afanasev:2020nur,Afanasev:2021fda}.
The only differences are that our expression is more general and that we do not take the limit of 
pointlike non-spreading target particle. This semiclassical assumption could be justified in atomic physics
but not in nuclear or particle physics where one usually deals with freely evolving wave packet collision.

Since all the delta-functions are absorbed in ${\cal I}$, we obtain a non-trivial distribution
in the full six-dimensional final phase space, which replaces the two-dimensional angular distribution
for the plane wave case \eqref{sigma-tot-dOmega}.
In particular, the total final momentum $\bP$ and the total final energy $E_f$ (or, alternatively,
the final system invariant mass $M_{\rm inv} = \sqrt{E_f^2 - P^2}$) are no longer fixed
and represent new dimensions for the kinematical analysis which were not available for the plane wave collisions.

Using the standard methods \cite{Landau4},
the expression for the Lorentz-invariant final phase space can be factorized 
into the phase space of the total motion and the relative motion of the two final particles:
\begin{equation}
\frac{d^3k_1'}{E_1'}\, \frac{d^3k_2'}{E_2'} = 
\frac{d^3k_1'}{E_1' E_2'}\, d^3P =
\left(\frac{|\bk_1'| d\Omega_{1}\, dE_f}{E_f}\right)_{\rm cmf}\! d^3P
= \beta(M_{\rm inv}) \cdot \frac{1}{2}d\Omega_{1, {\rm cmf}}\, dM_{\rm inv} d^3P\,.
\label{final-phase-space}
\end{equation}
Here, the label cmf indicates that the quantities are to be evaluated in the center of motion frame,
which, in turn, depends on $\bP$. Since $(E_f)_{\rm cmf} = M_{\rm inv}$, we introduced here 
\begin{equation}
\beta(M_{\rm inv}) = \sqrt{1 - \frac{4M^2}{M_{\rm inv}^2}}\,, \label{betaM}
\end{equation}
the velocity of each final particle in the center of motion frame. 
Notice that the angular distribution $d\Omega_{1, {\rm cmf}}$ is also defined in the center of motion frame; 
it depends on $\bP$ and is not a universal factor.
Thus, the expression \eqref{final-phase-space} is not the most convenient one 
if we aim to study the six-dimensional final distributions.
However in situations where ${\cal M}$ depends on the final momenta only through $P^\mu = (E_f, \bP)$,
the differential cross section, evaluated at fixed $E_f, \bP$, does not depend on the angles of $\bk_1'$.
Then we can perform the integration $\int d\Omega_{1, {\rm cmf}} = 4\pi$, divide the result by 2
due to the two identical particles in the final state,
and arrive at the cross section which is differential only with respect to the total final energy-momentum:
\begin{equation}
d\sigma = \frac{\pi^3}{L} \, \beta(M_{\rm inv}) \, |{\cal I}|^2 \, dM_{\rm inv} d^3P\,.
\label{dsigma-WP-int}
\end{equation}

\subsection{Time evolution and duration of the collision event}

Localized wave packets cannot be monochromatic and, therefore, they evolve in time.
The time dependence of $\psi(\br, t)$ arises from Eq.~\eqref{psi-def} due to the momentum dependence of the energy.
Exact expressions for $\psi(\br, t)$ for Gaussian and LG wave packets, both in the paraxial approximation and beyond, 
were already presented in \cite{Karlovets:2018iww,Karlovets:2020odl}.
In the present paper, we work in the paraxial approximation
and assume that the typical transverse momenta are much smaller than the average longitudinal momentum:
$1/\sigma_{1\perp} \ll p_{1z}$ and $1/\sigma_{2\perp} \ll |p_{2z}|$. 
By adapting the formalism of \cite{Karlovets:2020odl}
to the case when $\langle \bk_i\rangle = (0,0,p_{iz})$, we define $\varepsilon_i = \sqrt{m^2 + p_{iz}^2}$ and expand the
energies $E(k_i)$ as
\begin{eqnarray}
E_i &=& \sqrt{m^2 + k_i^2} = \sqrt{\varepsilon_i^2 + 2p_{iz}(k_{iz}-p_{iz}) + (k_{iz}-p_{iz})^2 + k_{i\perp}^2}\nonumber\\
&\approx& \varepsilon_i + v_i(k_{iz}-p_{iz}) + \frac{1}{2\varepsilon_i}\left[\frac{(k_{iz}-p_{iz})^2}{\gamma_i^2} + k_{i\perp}^2\right]\,.
\label{E-expansion}
\end{eqnarray}
Substituting these expressions into Eq.~\eqref{psi-def} and making use of the integrals
\begin{eqnarray}
\int_0^{2\pi} d\phi_k e^{i \ell \phi_k} e^{i r_\perp k_\perp \cos{(\phi_k - \phi_r)}} &=& 
2\pi i^\ell e^{i \ell \phi_r} J_\ell (r_\perp k_\perp),\label{integral-1}\\
\int_0^{\infty} k_\perp d k_\perp k_\perp^l e^{-\alpha k_\perp^2} J_\ell(\beta k_\perp)
	&=& \frac{\beta^\ell}{(2\alpha)^{\ell+1}} \exp\left(-\frac{\beta^2}{4\alpha}\right)\,,
	\qquad \mbox{Eq.~(6.631.4) of Ref.~\cite{integrals},}\label{integral-2}
\end{eqnarray}
we obtain the following coordinate space wave function for the LG state: 
\begin{equation}
\psi_1(\br, t) = \pi^{-3/4} \frac{i^\ell}{\sqrt{\ell!}}\frac{\sqrt{\sigma_{1\perp}^2\sigma_{1z}}}{\tilde\sigma_{1\perp}^2\tilde\sigma_{1z}}
\left(\frac{\sigma_{1\perp} r_\perp}{\tilde\sigma_{1\perp}^2}\right)^\ell e^{i\ell\phi_r}
\exp\left[-\frac{r_\perp^2}{2\tilde\sigma_{1\perp}^2}-\frac{(z-v_1 t)^2}{2\tilde\sigma_{1z}^2}\right]\, e^{-i\varepsilon_1 t + i p_{1z}z}\,.
\end{equation}
Here, we used the shorthand notation for the complex time-dependent combinations
\begin{equation}
\tilde\sigma_{1\perp}^2 = \sigma_{1\perp}^2 + i \frac{t}{\varepsilon_1}\,, \quad 
\tilde\sigma_{1z}^2 = \sigma_{1z}^2 + i \frac{t}{\gamma_1^2 \varepsilon_1}\,.\label{tilde-sigma-1}
\end{equation}
The probability density then takes the following form
\begin{equation}
|\psi_1(\br, t)|^2 = \frac{1}{\pi^{3/2}\ell!} \, \frac{1}{\sigma_{1\perp}^2(t) \sigma_{1z}^2(t)}
\left(\frac{r_\perp^2}{\sigma_{1\perp}^2(t)}\right)^\ell\exp\left[-\frac{r_\perp^2}{\sigma_{1\perp}^2(t)}
-\frac{(z-v_1 t)^2}{\sigma_{1z}^2(t)}\right]\,, 
\end{equation} 
where the effective time-dependent localization lengths are
\begin{equation}
\sigma_{1\perp}^2(t) \equiv \sigma_{1\perp}^2 \left(1 + \frac{t^2}{\sigma_{1\perp}^4\varepsilon_1^2}\right)\,,\quad
\sigma_{1z}^2(t) \equiv \sigma_{1z}^2 \left(1 + \frac{t^2}{\sigma_{1z}^4\gamma_1^4 \varepsilon_1^2}\right)\,.\label{sigma-perp-z-t}
\end{equation}
We recovered the well-known spreading of the wave packet as it propagates,
with the typical spreading time being $\sigma_{1\perp}^2\varepsilon_1$ for the transverse spreading 
and $\sigma_{1z}^2\gamma_1^2\varepsilon_1$ for the longitudinal one.
For a wave packet with $\sigma_{1z} = \sigma_{1\perp}/\gamma_1$, the two brackets in Eqs.~\eqref{sigma-perp-z-t} are equal, 
and the spreading wave packet preserves its shape. 

For the Gaussian state, we observe a similar dynamics. Taking into account all its shifts, we get
\begin{equation}
\psi_2(\br, t) = \pi^{-3/4} \frac{\sqrt{\sigma_{2\perp}^2\sigma_{2z}}}{\tilde\sigma_{2\perp}^2\tilde\sigma_{2z}}
\exp\left[-\frac{(\br_\perp-\bb_\perp)^2}{2\tilde\sigma_{2\perp}^2}
-\frac{(z - b_z - v_2 (t-\tau))^2}{2\tilde\sigma_{2z}^2}\right]\, e^{-i\varepsilon_2 t + i p_{2z}z}\,.
\end{equation}
with 
\begin{equation}
\tilde\sigma_{2\perp}^2 = \sigma_{2\perp}^2 + i \frac{t-\tau}{\varepsilon_2}\,, \quad 
\tilde\sigma_{2z}^2 = \sigma_{2z}^2 + i \frac{t-\tau}{\gamma_2^2 \varepsilon_2}\,.\label{tilde-sigma-2}
\end{equation}
The only difference now is that the moment of maximal focusing is at $t=\tau$ and is located at $\bb$, not at the origin.

\subsection{Luminosity integral and the impulse approximation}

The explicit expression for the coordinate wave functions allows us to calculate the luminosity $L$ in Eq.\eqref{lumi}.
Let us first evaluate it for collision of two Gaussian wave packets with a nonzero transverse shift $\bb_\perp$
but with $b_z = 0$, $\tau=0$.
After performing the transverse integration, we get
\begin{eqnarray}
L &=& \frac{|v_1 - v_2|}{\pi^2} \int dt\, dz\, 
\frac{\sigma^2_{1\perp}\sigma^2_{2\perp}}{\sigma^2_{1\perp}(t)\sigma^2_{2\perp}(t)[\sigma^2_{1\perp}(t) + \sigma^2_{2\perp}(t)]}
\exp\left[-\frac{b_\perp^2}{\sigma^2_{1\perp}(t) + \sigma^2_{2\perp}(t)}\right]\nonumber\\[2mm]
&&\qquad\times \ \frac{\sigma_{1z}\sigma_{2z}}{\sigma^2_{1z}(t)\sigma^2_{2z}(t)}
\exp\left[-\frac{(z-v_1t)^2}{\sigma^2_{1z}(t)}-\frac{(z-v_2t)^2}{\sigma^2_{2z}(t)}\right]\,.
\end{eqnarray}
Performing the $z$ integration leads to
\begin{eqnarray}
L &=& \frac{|v_1 - v_2|}{\pi^{3/2}} \int dt\, 
\frac{\sigma^2_{1\perp}\sigma^2_{2\perp}}{\sigma^2_{1\perp}(t)\sigma^2_{2\perp}(t)[\sigma^2_{1\perp}(t) + \sigma^2_{2\perp}(t)]}
\left[\frac{\sigma^2_{1z}\sigma^2_{2z}}{\sigma^2_{1z}(t)\sigma^2_{2z}(t)[\sigma^2_{1z}(t) + \sigma^2_{2z}(t)]}\right]^{1/2}\nonumber\\[2mm]
&&\qquad\times \exp\left[-\frac{b_\perp^2}{\sigma^2_{1\perp}(t) + \sigma^2_{2\perp}(t)}\right]
\exp\left[-\frac{t^2(v_1-v_2)^2}{\sigma^2_{1z}(t) + \sigma^2_{2z}(t)}\right]\,.\label{L-int-3}
\end{eqnarray}
Instead of evaluating the $t$ integral exactly, we notice that the main dependence comes from the last exponential,
which drops significantly over the timescale $t_c = \sqrt{\sigma_{1z}^2+ \sigma_{2z}^2}/|v_1-v_2|$,
which we call duration of the collision event.
The crucial step is to assume that during collision the longitudinal and transverse localization scales 
\eqref{tilde-sigma-1} and \eqref{tilde-sigma-2}
do not change significantly.
This condition is satisfied automatically for the longitudinal scale due to
$\sigma_{iz}p_{iz} \gg 1$, 
so that the main constraint comes in the form of an upper limit on $\sigma_{1z}^2 + \sigma_{2z}^2$:
\begin{equation}
\sqrt{\sigma_{1z}^2 + \sigma_{2z}^2} < \sigma_{i\perp}^2 \varepsilon_i |v_1-v_2| \,.\label{impulse-conditions}
\end{equation}
We call this assumption the {\em impulse approximation}.
Put simply, it means that the wave packets cannot be too long for a given transverse localization scale.
In our previous paper \cite{Ivanov:2022sco}, we used Bessel beams of infinite longitudinal extent,
and, in order to limit the collision time, we had to introduce an artificial regularization procedure.
Now, with sufficiently compact wave packets, we get rid of this auxiliary procedure.

Within the impulse approximation, the remaining time integral can be evaluated in a straightforward way 
by replacing all the transverse and longitudinal parameters $\sigma^2(t) \to \sigma^2(0) = \sigma^2$.
Notice that, at very large $t$, the last exponential factor does not vanish and approaches a finite but exponentially suppressed value. 
However the integral still converges at large $t$ thanks to the $t$-dependent prefactors. 
Thus, the contribution from the late-time tails remains exponentially suppressed, and we obtain
\begin{equation}
L = \frac{1}{\pi}\frac{1}{\sigma^2_{1\perp} + \sigma^2_{2\perp}} \exp\left(-\frac{b_\perp^2}{\sigma^2_{1\perp} + \sigma^2_{2\perp}}\right)\,,
\label{L-1}
\end{equation}
which coincides with Eq.~(52) of \cite{Karlovets:2020odl}. 
However when deriving it, we did not need to assume that the colliding particles were ultrarelativistic.

Turning to the luminosity integral for the collision of LG and Gaussian wave packets,
we notice that the longitudinal and time integrations can be done exactly in the same way,
so that the luminosity can be expressed as 
\begin{equation}
L = \frac{1}{\pi^2 \ell!} \frac{1}{\sigma_{1\perp}^2\sigma_{2\perp}^2} \int d^2 r_\perp \left(\frac{r_\perp^2}{\sigma_{1\perp}^2}\right)^\ell
\exp\left[-\frac{r_\perp^2}{\sigma_{1\perp}^2}-\frac{(\br_\perp-\bb_\perp)^2}{\sigma_{2\perp}^2}\right]\,.
\end{equation}
This integral can be taken exactly and expressed compactly via the Laguerre polynomials:
\begin{equation}
\label{luminosity-result}
	L= \frac{1}{\pi} \frac{1}{\sigma_{1\perp}^2+\sigma_{2\perp}^2} 
	\left( \frac{\sigma_{2\perp}^2}{\sigma_{1\perp}^2+\sigma_{2\perp}^2}\right)^{\ell}
	\exp{\left( - \frac{b_\perp^2}{\sigma_{1\perp}^2 + \sigma_{2\perp}^2}\right)} \,
	L_\ell\left( -\frac{b_\perp^2}{\sigma_{2\perp}^2} 
	\frac{\sigma_{1\perp}^2}{\sigma_{1\perp}^2+\sigma_{2\perp}^2}\right)\,.   
\end{equation}

Let us now take into account the non-zero $b_z$ and $\tau$.
Each wave packet reaches its minimal transverse and longitudinal size at different $z$ positions and at different instants.
However the collision duration is still short, and during this overlap the localization scales remain almost frozen. 
Therefore, within the impulse approximation, one can re-use the expression \eqref{L-int-3}
but replace all $\sigma^2(t)$ with their values at the moment of collision.

Thus, if we stick to the impulse approximation, the dependence of all the expressions 
on $b_z$ and $\tau$ becomes redundant: these two parameters only affect the values of $\sigma_z$
and $\sigma_\perp$ at the moment of collision. From now on, we will use just 
$\sigma_z$ and $\sigma_\perp$ (at the moment of collision) 
instead of their exact $t$-dependent expressions.
The parameter $b$ will refer to the transverse impact parameter only $b \equiv |\bb_\perp|$.

\section{Threshold behavior and the superkick regime}\label{section-main-results}

\subsection{The cross section far from the threshold}\label{subsection-far-from-threshold}

Having defined the impulse approximation and obtained the luminosity integrals for the LG vs. Gaussian wave packet scattering,
we can now explore the cross section in various regimes.
We are particularly interested in reproducing the superkick effect within the above formalism
and its possible influence on the threshold behavior proposed in \cite{Afanasev:2020nur,Afanasev:2021fda}.

As a first cross-check of our formalism, we consider the same process $\phi\phi \to \Phi\Phi$
sufficiently far above the energy threshold and compare the results with the plane wave cross section \eqref{sigma-tot-dOmega}.
For simplicity, we now assume that the tree-level invariant amplitude ${\cal M}$ is constant
and switch to the center of motion reference frame defined in terms of the average momenta of the colliding wave packets:
$p_{1z} + p_{2z} = 0$, so that $p_{1z} = (0,0,p_0)$, $p_{2z} = (0,0,-p_0)$.
The average energy of each particle in this frame is denoted as $E_0 = \sqrt{m^2+p_0^2}$;
the absolute value of the initial particles velocity is $v_0 = p_0/E_0$.

``Far above the threshold'' means that the momenta of the final particles $|\bk'_\perp|$ are much larger than
$|\bP_\perp|$ and $P_z$, but it does not imply that the final particles are ultrarelativistic.
Then, in the final phase space \eqref{final-phase-space}, 
the $\bP$-dependent angular measure $d\Omega_{1, {\rm cmf}}$ can be approximated by a universal $d\Omega_1$,
the same as we had for the plane-wave cross section. At the same time, $\beta(M_{\rm inv})$ can be approximated by $\beta$
and $d M_{\rm inv} d^3P$ can be replaced by $d^4P$.
Thus, the cross section \eqref{dsigma-WP} becomes
\begin{equation}
d\sigma = \frac{\pi^2}{2 L} \, \beta \, d\Omega_1\, |{\cal I}|^2\,  d^4P\,.
\label{dsigma-WP-2}
\end{equation}
Now, in order to compare it with the plane wave cross section, we keep the single-particle angular distribution
and integrate over the total final system motion: $\int |{\cal I}|^2\,  d^4P$.
To perform it without dealing with explicit wave functions, we return to the definition of ${\cal I}$ 
given in Eq.~\eqref{cal-I} and transform the four-dimensional delta function as
\begin{equation}
\delta^{(4)}(k_1+k_2-P) = \frac{1}{(2\pi)^4}\int d^4x\, e^{ik_1x + ik_2x - iPx}\,.\label{delta4}
\end{equation}
Using this representation twice in $|{\cal I}|^2$, we can perform all the $k_i$ integrations and 
express the result as
\begin{equation}
\int |{\cal I}|^2 d^4P = \frac{|{\cal M}|^2}{(2\pi)^4\, 4E_0^2} \int d^4x\, |\psi_1(\br,t)|^2 |\psi_2(\br,t)|^2\,.
\label{I2d4P}
\end{equation}
The remaining integral is exactly the luminosity function \eqref{lumi} without the relative velocity: $L/|v_1-v_2| = L/(2v_0)$.
Thus, we arrive at the following differential cross section
\begin{equation}
d\sigma = \frac{|{\cal M}|^2}{256\pi^2 E_0 p_0}\, \beta\, d\Omega\,,\label{dsigma-like-PW}
\end{equation}
which coincides with the plane wave cross section \eqref{sigma-tot-dOmega}.

This result is expected, especially after the systematic study \cite{Karlovets:2020odl},
but it is nevertheless remarkable. We arrived at the plane-wave cross section {\em without taking the plane wave limit}.
This conclusion hinges upon three assumptions: the paraxial approximation is applicable,
the invariant amplitude is constant, and the process occurs sufficiently far above the threshold.
With these assumptions, the result is valid for {\em arbitrarily shaped wave packets},
including the LG and Gaussian collision in the superkick kinematics. 
The cross section stays the same regardless whether the superkick effect is present or not:
for $b = 0$ (no superkick),
$\sigma_{2\perp} < b \ll \sigma_{1\perp}$ (superkick expected),
and $b \gg \sigma_{1\perp}$ (no superkick again). 
In these three situations, the luminosity functions $L$ and the event rates $dW$ will certainly be different,
but their ratio, the $P$-integrated cross section, does not depend on $b$.

\subsection{Reproducing the superkick}\label{subsection-superkick}

The fact that the $P$-integrated cross section well above the threshold is $\bb_\perp$-independent 
does not mean that the differential distribution is equally insensitive to $\bb_\perp$.
It certainly does depend on $\bb_\perp$ and displays an azimuthal bias which is the origin of the superkick effect itself.
Within the same kinematical assumption (process is well above the threshold), 
one can compute the average total transverse momentum:
$\lr{\bP_\perp} = \int \bP_\perp\, d\sigma/\int d\sigma$, and track its $\bb_\perp$ dependence.
The explicit evaluation of ${\cal I}$ described in Appendix~\ref{appendix-I} shows that, for the constant invariant amplitude,
the expression for $|{\cal I}|^2$ factorizes into the longitudinal and the transverse parts.
As a result, 
\begin{equation}
\label{average-P-perp}
	\lr{\bP_\perp} = \frac{\int d^2 P_\perp \bP_\perp |\mathcal{I}_\perp|^2}{\int d^2 P_\perp |\mathcal{I}_\perp|^2}.
\end{equation}
This result serves as a confirmation of the validity of the regularization procedure used in our previous work \cite{Ivanov:2022sco}.
With the explicit expressions for these integrals,
we get a compact analytical expression, in terms of Laguerre polynomials, for the final transverse momentum for a given $\bb_\perp = b\, \bbe_x$:
\begin{equation} 
\label{final-result-PT}
	\lr{\bP_\perp} = \bbe_y\frac{b}{\sigma_{2\perp}^2}
	 \frac{\displaystyle L_{\ell-1}^{1}\left(- \frac{b^2}{\sigma_{2\perp}^2} \frac{\sigma_{1\perp}^2}{\sigma_{1\perp}^2 + \sigma_{2\perp}^2}\right)}
	 {\displaystyle L_{\ell}\left(- \frac{b^2}{\sigma_{2\perp}^2} \frac{\sigma_{1\perp}^2}{\sigma_{1\perp}^2 + \sigma_{2\perp}^2}\right)}
	=  \frac{\bbe_y}{2} \frac{\sigma_{1\perp}^2 + \sigma_{2\perp}^2}{\sigma_{1\perp}^2}\,  \partial_{b}\, 
	\ln{ \left[ L_{\ell}\left(- \frac{b^2}{\sigma_{2\perp}^2} \frac{\sigma_{1\perp}^2}{\sigma_{1\perp}^2 + \sigma_{2\perp}^2}\right) \right] }\,.
\end{equation}
If the Gaussian wave packet is much more compact than the LG beam, $\sigma_{2\perp} \ll \sigma_{1\perp}$, 
the factor $(\sigma_{1\perp}^2 + \sigma_{2\perp}^2)/\sigma_{1\perp}^2 \to 1$,
and the result \eqref{final-result-PT} coincides with Eq.~(3.7) of \cite{barnett2013superkick}, 
where a non-relativistic probe atom was considered in the semiclassical field of an optical vortex.

\begin{figure}[!h]
	\centering
	\includegraphics[width=0.6\textwidth]{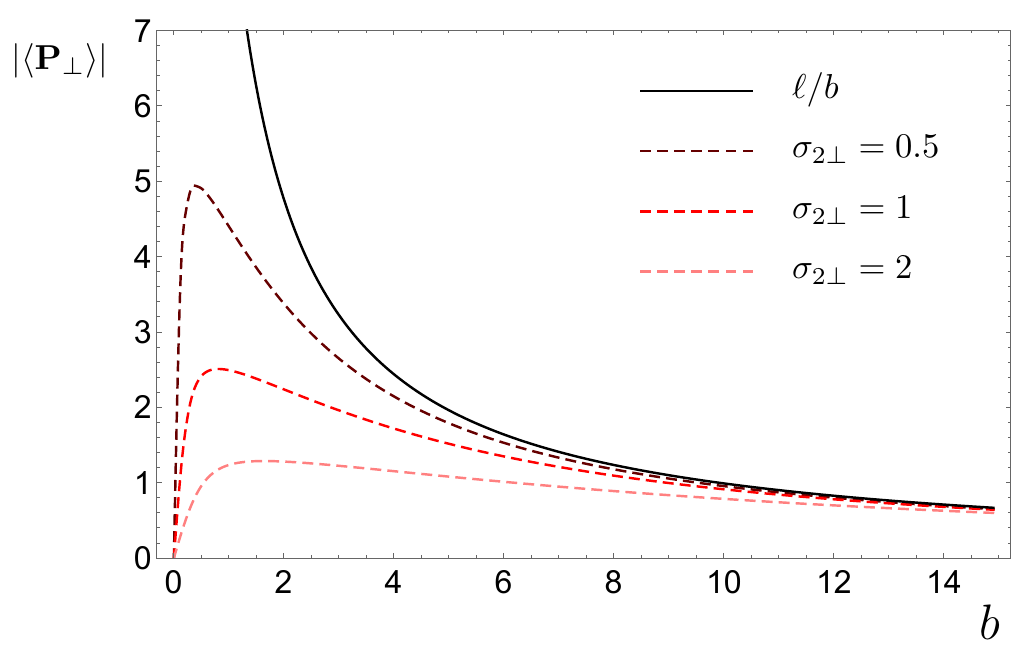}
	\caption{The total transverse momentum $|\lr{\bP_\perp}|$ in the $\ell =10$ LG vs. Gaussian wave packet collision
	as a function of the impact parameter $b$.
	The kinematical parameters are expressed in units of an arbitrary common energy scale and are given in Eq.~\eqref{parameters-superkick}.}
	\label{fig-superkick}
\end{figure}

In Fig.~\ref{fig-superkick} we plot $|\lr{\bP_\perp}|$ as a function of $b$ for $\ell=10$ and for several values of 
$\sigma_{2\perp}$. 
Since we consider a toy model, not a specific particle collision process, we introduce here
an arbitrary but fixed energy scale $\Lambda$ and express all dimensional parameters
in units of $\Lambda$ (for masses, energies, momenta) or $\Lambda^{-1}$ (for $\sigma_{iz}$, $\sigma_{i\perp}$, $b$).
In these units, the dimensional parameters used for the plot in Fig.~\ref{fig-superkick} are
\begin{equation}
m = 10, \quad M = 100, \quad \sigma_{1\perp}=10,\quad \sigma_{1z} = 40\,,\quad
\sigma_{2\perp}=\{0.5,\ 1,\ 2\},\, \quad \sigma_{2z}=4\,.
\label{parameters-superkick}
\end{equation}
The curves agree with our results presented in \cite{Ivanov:2022sco},
with the Bessel beam parameter $\varkappa$ replaced by the inverse scale $1/\sigma_{1\perp}$.
The superkick corresponds to the final momentum being much larger than the typical transverse momentum
of the LG state, which is $1/\sigma_{1\perp} = 0.1$.
It is indeed observed within the entire region shown.
However, the semiclassical expectation $|\lr{\bP_\perp}| = \ell/b$ holds only within 
the region given by Eq.~\eqref{scales}. 
For example, at $b = \sigma_{2\perp}$, the effect is about four times weaker than given by the $\ell/b$ curve.
Thus, we reproduce the superkick effect and its main dependences with the LG vs. Gaussian wave packet scattering.

Additional insights in the superkick effect can be obtained through the azimuthal distribution of the cross section.
One expects that the average $\lr{\bP_\perp}$ is orthogonal to $\bb$, but the exact $\bP_\perp$ distribution,
as well as the event-by-event correlations between $\bP_\perp$ and, say, $\bk'_{1\perp}$
may contains additional features which are sensitive to the scattering process.
Since we work in the toy model wit scalar particles, we do not investigate these details.
However in a future study of the superkick effect in realistic scattering, such as the Compton scattering,
one should pay attention to this distribution.
Examples of non-trivial azimuthal distribution in the vortex Compton scattering, even for $b = 0$, 
can be found in \cite{Maruyama:2017ptl}. For a non-zero impact parameter,
we expect the azimuthal distribution to be a richer source of information.

\subsection{The energy behavior near the threshold}\label{subsection-near}

Near the threshold, we have $|\bk'_\perp| \sim |\bP_\perp|$, so that $\beta(M_{\rm inv}) \ll 1$ and it cannot be approximated by a constant.
From the above discussion, we can conclude that it is the $P$-integral of $\beta(M_{\rm inv})$ which 
governs the near-threshold cross section behavior. 
Thus, we define the key quantity which we need to explore:
\begin{equation}
\tilde \beta = 	\frac{\int \beta(M_{\rm inv}) \, d M_{\rm inv}\,  |\mathcal{I}|^2 \, d^3P }{\int |\mathcal{I}|^2\, d^4 P}\,.
\label{tilde-beta}
\end{equation}
This function depends on the initial particle average energies $\varepsilon_1$ and $\varepsilon_2$,
or on $E_0$ if we evaluate it in the average center of motion reference frame.

For the following analysis, it is convenient to introduce the following combinations:
\begin{equation}
\Sigma^2_\perp = \frac{\sigma_{1\perp}^2 \sigma_{2\perp}^2}{\sigma_{1\perp}^2+\sigma_{2\perp}^2} \,,\quad
\Sigma^2_z = \frac{\sigma_{1z}^2 \sigma_{2z}^2}{\sigma_{1z}^2+\sigma_{2z}^2} \,.\label{Sigmas}
\end{equation}
In the limit $\sigma_1 \gg \sigma_2$, they approach the smaller localization scales:
$\Sigma_\perp \approx \sigma_{2\perp}$,  $\Sigma_z \approx \sigma_{2z}$.
In Appendix~\ref{appendix-I}, we give the details of the explicit evaluation of ${\cal I}$ in the paraxial approximation.
In the average center of motion frame, the result for its modulus squared is
\begin{eqnarray} \label{cal-I-2}
	|\mathcal{I}|^2 &=& |\mathcal{M}|^2 \frac{2}{(2\pi)^7 \ell!} \frac{\sigma_{1z} \sigma_{2z}}{4 E_0^2 v_0^2} \cdot 
	\left( \frac{\Sigma_{\perp}^2}{\sigma_{1\perp}^2 + \sigma_{2\perp}^2} \right)^{\ell+1}   
	\left[ \frac{b^2}{\sigma_{2\perp}^2} + P_\perp^2 \sigma_{2\perp}^2 +2 b P_\perp \sin{(\varphi_P - \varphi_b)} \right]^\ell 
	\nonumber \\[2mm]
	&& \times 
	\exp\left[  - \frac{b^2}{\sigma_{1\perp}^2+\sigma_{2\perp}^2} 
	- P_\perp^2 \Sigma_{\perp}^2
	- \frac{(\delta E + P_z v_0)^2 \sigma_{1z}^2 + (\delta E - P_z v_0)^2 \sigma_{2z}^2}{4v_0^2} \right].
\end{eqnarray} 
Here, $\delta E \equiv E_f- 2E_0$.
We have checked that $\int |{\cal I}|^2 d^4P$ coincides with the result \eqref{I2d4P}.

Before passing to numerical results, let us get some feeling of the sub-threshold behavior of the function $\tilde \beta$.
Due to $M_{\rm inv} \ge 2M$, we get the lower limit on the total final energy:
\begin{equation}
E_f = \sqrt{M_{\rm inv}^2 + P^2} \ge \sqrt{4M^2 + P^2} \ge 2M\,.\label{Ef-limit}
\end{equation}
Suppose that $E_0 < M$. Then $\delta E = E_f- 2E_0 > 2(M-E_0)$. Substituting $(\delta E)_{\rm min} = 2(M-E_0)$ into the 
longitudinal exponential of Eq.~\eqref{cal-I-2} and selecting the value of $P_z$ which minimizes the exponential in the suppression factor, 
we obtain the following exponential suppression:
\begin{equation}
\exp\left[- \frac{(\delta E)_{\rm min}^2 \Sigma^2_{z}}{v_0^2}\right]=
\exp\left[- \frac{4(M-E_0)^2 \Sigma^2_{z}}{v_0^2}\right]\,.
\label{energy-suppression}
\end{equation}
Thus, for $\sigma_{1z} \gg \sigma_{2z}$, the cross section
can extend below the threshold by the amount $|E_0 - M| \approx v_0/(2\sigma_{2z})$.

\begin{figure}[!h]
	\centering
	\includegraphics[width=0.5\textwidth]{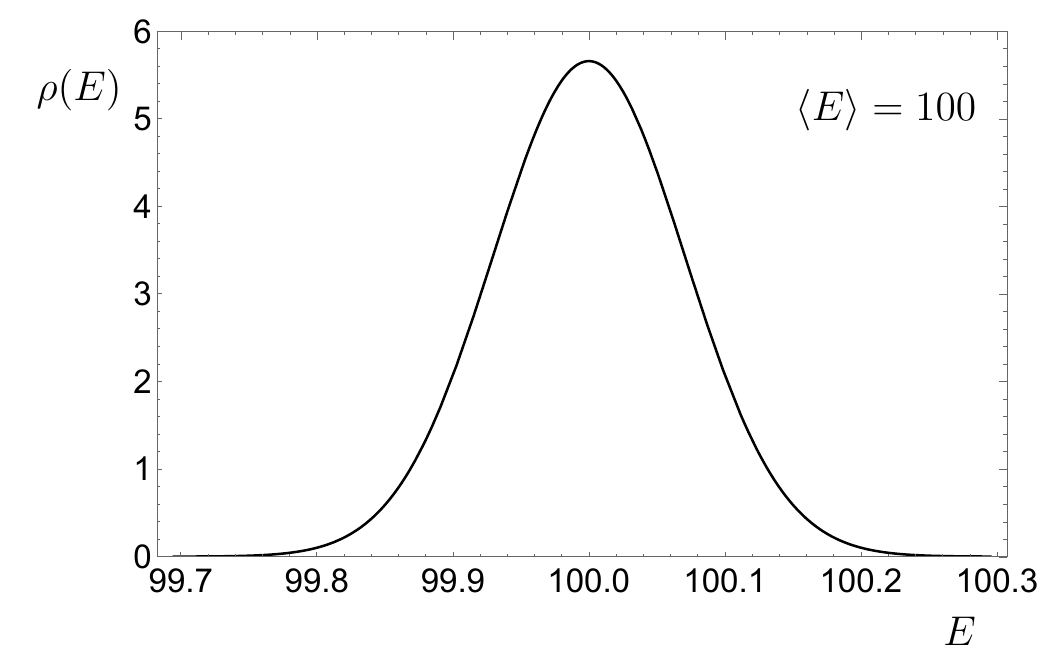}
	\caption{The energy distribution $\rho(E)$ 
	of a Gaussian wave packet with $m=10$, $\lr{E} = 100$, $1/\sigma_{1\perp} = 1$,  $1/\sigma_{1z} = 0.1$,
	which lead to the parameter $E_0 \approx 99.995$.
	All quantities are expressed in units of an arbitrary common energy scale.}
	\label{fig-energy-distribution}
\end{figure}

We stress that this apparent sub-threshold behavior does not contradict energy conservation.
It is just a manifestation of the simple fact that the initial wave packets are not 
monochromatic but have energy distribution within the range of the order of $v_0 \delta k_{iz} \approx v_0/\sigma_{iz}$.
To make this point clear, we show in Fig.~\ref{fig-energy-distribution}, the energy distribution $\rho(E)$ 
of a compact ultrarelativistic Gaussian wavepacket, normalized by $\int \rho(E)dE = 1$. This example corresponds to
$m=10$, $\lr{E} = 100$, $1/\sigma_{\perp} = 1$,  $1/\sigma_{z} = 0.1$, all expressed in a common energy unit.
As expected, the width of this distribution is of the order of $v_0/\sigma_{z} \approx 0.1$.
At the same time, the difference between the parameter $E_0 = \sqrt{m^2+\lr{\bk}^2}$ 
and the true average energy in this state $\lr{E} = \lr{E(\bk)}$, where $E(\bk) = \sqrt{m^2+\bk^2}$,
is tiny, of the order of $\bk_\perp^2/2E_0 \approx 1/(2\sigma_{\perp}^2 E_0) \approx 0.005$.
Thus, the shift of the average energy due to transverse motion inside the wave packet is negligible.

Consequently, a non-zero cross section in the nominally ``sub-threshold'' region
does not mean, of course, that the true threshold shifts to lower energies.
It is just the manifestation of the fact that an ultrarelativistic wave packet
contains plane-wave components above the threshold $E>M$ even if the wave packet energy, on average, is sub-threshold.

Let us now track the main effect of a non-zero $b$ on this threshold smearing.
The sensitivity to $\bb_\perp$ appears through the $\bP_\perp$ integral; 
thus, we need to take a closer look at the $\bP_\perp$ dependence 
of the integral in the numerator of Eq.~\eqref{tilde-beta}.
At non-zero $\bP_\perp$, the lower limit on $\delta E$ increases by $P_\perp^2/2M$.
Thus, in addition to the sub-threshold energy suppression factor \eqref{energy-suppression},
we obtain an extra suppressing factor
\begin{equation}
\exp\left[- \frac{(M-E_0)}{M} P_\perp^2\Sigma^2_{z}\right]\,.
\label{Pperp-suppression}
\end{equation}
Effectively it amount to the replacement
\begin{equation}
\Sigma_{\perp}^2 \to \Sigma_{\perp}^2 + \Delta \Sigma_\perp^2\,,
\quad \mbox{where}\quad
\Delta \Sigma_\perp^2 = \frac{(M-E_0)}{M} \Sigma^2_{z}\,.
\end{equation}
It is this change of $\Sigma_{\perp}^2$ which, after the $\bP_\perp$ integration, leads to the $b$-dependence of the results.
For example, for $\ell = 1$ we expect the following modification of the sub-threshold behavior of $\tilde\beta$
compared to the Gaussian-Gaussian case:
\begin{equation}
\tilde\beta \propto \frac{\Sigma_{\perp}^2}{\Sigma_{\perp}^2 + \Delta \Sigma_\perp^2}
\frac{\sigma_{2\perp}^4 + b^2 (\Sigma_{\perp}^2 + \Delta \Sigma_\perp^2)}{\sigma_{2\perp}^4 + b^2\Sigma_{\perp}^2}\,.
\end{equation}
As $b$ grows, we observe a smooth transition of $\tilde\beta$
from $\Sigma_{\perp}^2/(\Sigma_{\perp}^2 + \Delta \Sigma_\perp^2)$ at $b=0$ to 1 at large $b$.
That is, for the LG vs Gaussian sub-threshold scattering, we expect an additional suppression,
not an enhancement, with respect to the two Gaussian state collisions.
This suppression is the strongest at $b=0$ and becomes weaker as $b$ grows.

\subsection{Numerical results}

\begin{figure}[!h]
	\centering
	\includegraphics[width=0.5\textwidth]{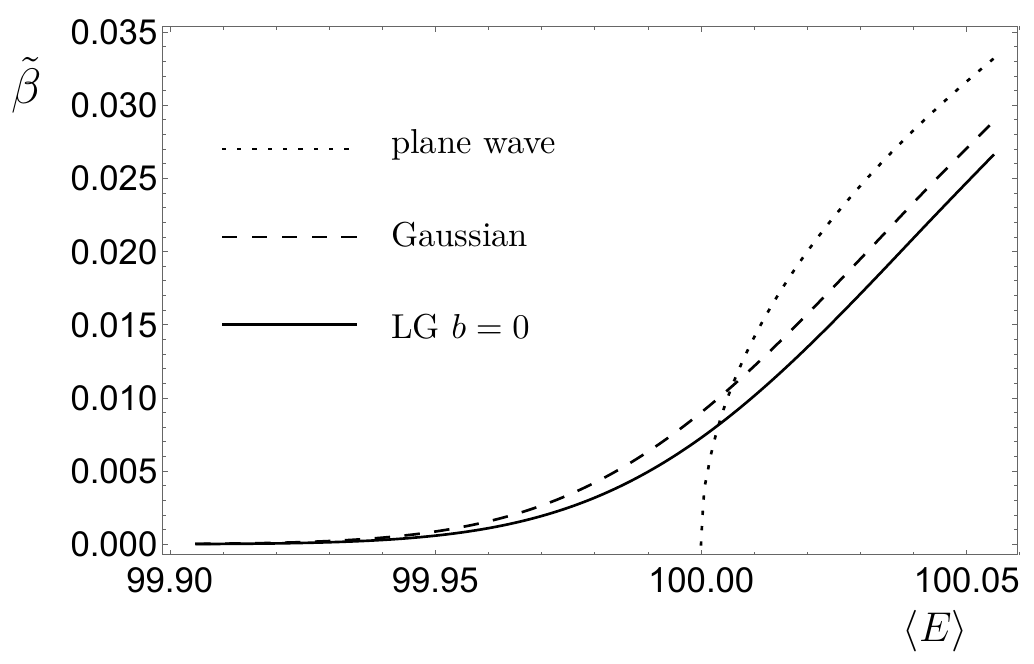}
	\caption{Energy dependence of the function $\tilde\beta$ near the threshold.
	Shown are the plane wave case (dotted line) and two wave packet collision configurations:
	Gaussian vs Gaussian (dashed line) and LG ($\ell = 10$) vs Gaussian 
	at $b = 0$. The other kinematical parameters are as in Eq.~\eqref{parameters}. 
	All quantities are expressed in units of an arbitrary common energy scale.}
	\label{fig-threshold-1}
\end{figure}

\begin{figure}[!h]
	\centering
	\includegraphics[width=0.45\textwidth]{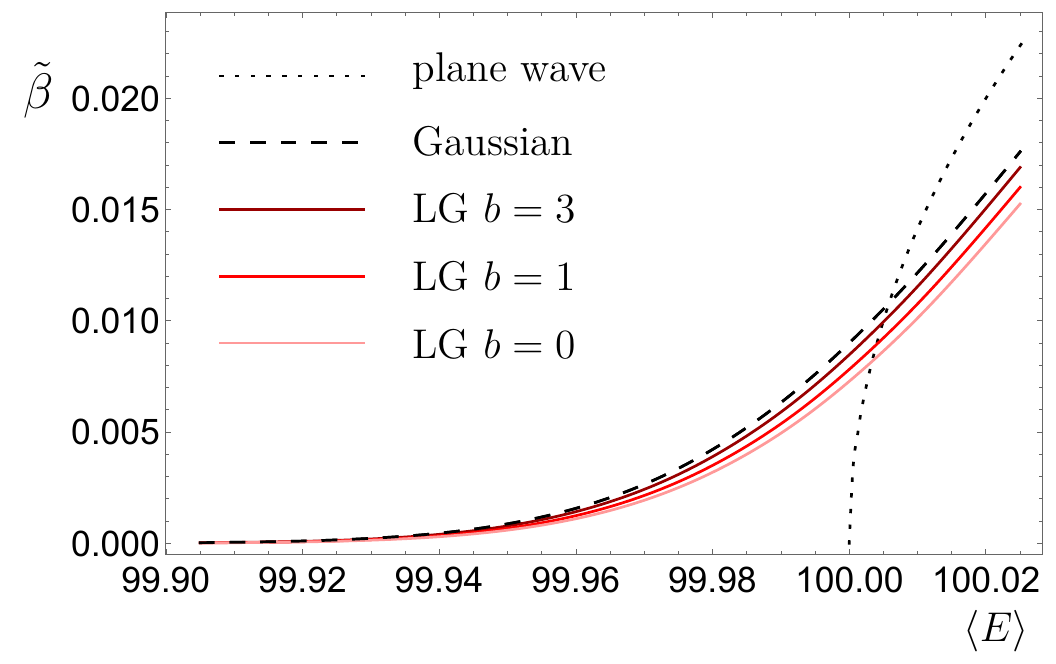}
	\hspace{5mm}
	\includegraphics[width=0.45\textwidth]{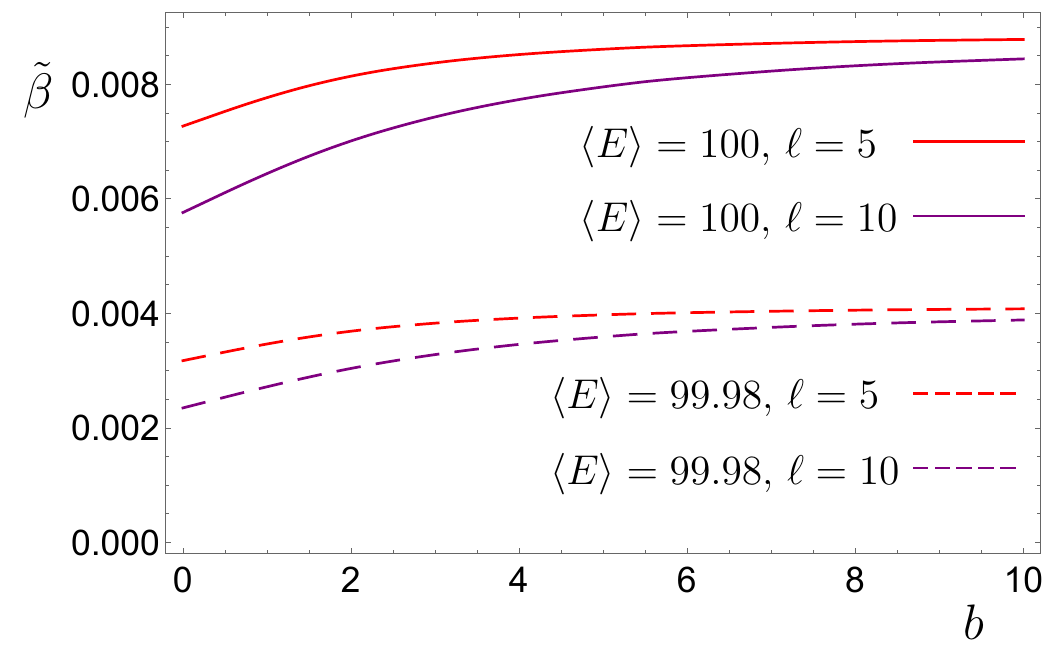}
	\caption{Left: The energy dependence of $\tilde\beta$ for the LG vs Gaussian collision with $\ell = 10$
	and different impact parameters $b$. The Gaussian-Gaussian case (dashed line) is $b$-independent.
	Right: The $b$-dependence of $\tilde\beta$ at two fixed energies and for two option of $\ell$.
	The kinematical parameters are as in Eq.~\eqref{parameters}.}
	\label{fig-threshold-2}
\end{figure}

In Figs.~\ref{fig-threshold-1} and \ref{fig-threshold-2} we show numerical results for $\tilde\beta$
defined in Eq.~\eqref{tilde-beta} as a function of energy in the threshold region;
the cross section is proportional to $\tilde\beta$ in the vicinity of the threshold.
In our calculations, we chose parameters which put in evidence the $b$-dependence of the threshold smearing.
Expressing the dimensional parameters in units of an arbitrary but fixed (inverse) mass scale, we select
\begin{equation}
m = 10, \quad M = 100, \quad \sigma_{1\perp}=10,\quad \sigma_{2\perp}=1,\quad \sigma_{iz}=10\, \sigma_{i\perp}\,.
\label{parameters}
\end{equation}
These values are well compatible with the paraxial approximation as well as
with the criterion for validity of the impulse approximation \eqref{impulse-conditions}.
As the energy variable, we choose not the input parameter $E_0$, which may seem artificial, 
but $\lr{E} \equiv \lr{E(\bk_2)}$, the true average energy of the compact Gaussian wave packet.
As mentioned above, that the difference between $E_0$ and $\lr{E}$ is tiny,
of the order of $0.005$ in our example. 
The ``sub-threshold penetration'' region is $M-E \sim 0.05$, driven by the energy distribution
in the Gaussian wave packet.
Notice that we take $\sigma_{iz}$ to be much larger than $\sigma_{i\perp}$ instead of 
taking the small values $\sigma_{i\perp}/\bar\gamma_i$; 
this is done in order to expose the $b$-dependence of the threshold effects.

In Fig.~\ref{fig-threshold-1} we compare the energy behavior of $\tilde\beta$
for three cases: the plane wave final velocity $\beta$ as in Eq.~\eqref{sigma-tot-PW} 
(dotted line), the case of two Gaussian wave packets collisions (dashed line), 
which is calculated with the above formulas at $\ell = 0$,
and the LG wave packet with $\ell = 10$ colliding centrally with the compact Gaussian wave packet.
The sharp threshold behavior characteristic of the plane wave cross section is blurred 
once the finite width energy distribution for wave packet collisions is taken into account.
We see that the Gaussian-Gaussian cross section goes higher than the LG vs Gaussian, and the
suppression factor at $b = 0$ is indeed sizable.
We checked that sufficiently far above the threshold, at $\lr{E} - M \gg 1/\sigma_{1z}$, 
the three curves converge and approach their asymptotic value $\beta \to 1$. 
This confirms our finding in Section~\ref{subsection-far-from-threshold} 
of the identical cross sections above the threshold.

In Fig.~\ref{fig-threshold-2} we explore how the LG vs Gaussian cross section depends on the impact parameter $b$.
The left plot again displays the same quantity $\tilde\beta$ as a function of $\lr{E}$ but now 
it is plotted for several values of $b$. Confirming the above analytical results, we observe
that the LG vs Gaussian collision cross section is suppressed with respect to the Gaussian-Gaussian case
in a $b$-dependent manner. The strongest suppression is at $b = 0$ and it gradually weakens as $b$ grows.
We checked that at $b > 10\sigma_{2\perp}$, when the compact Gaussian wave packet is no longer probing the phase vortex,
the curve is already indistinguishable from the Gaussian-Gaussian case.

This dependence is made explicit in the right plot of Fig.~\ref{fig-threshold-2}
where we show how $\tilde\beta$ depends on $b$.
The two pairs of curves correspond to $\ell = 5$ and $\ell = 10$
computed at $\lr{E} = 100$ (the upper pair) and $\lr{E} = 99.98$ (the lower pair).
The small-$b$ dips in these curves become more pronounced for larger $\ell$.
We also checked that the magnitude of the dip depends on the value of $\sigma_{2\perp}$:
the more compact the Gaussian probe wave packet, the stronger the suppression at $b \to 0$.

The physics picture which emerges from our results is quite different from the conclusions 
of \cite{Afanasev:2020nur,Afanasev:2021fda}.
In those works, the photoinduced hadronic processes $\gamma d \to pn$ and $\gamma p \to \Delta$
with vortex gamma photons and localized target probes
were predicted to exhibit a significant {\em increase} of the cross section as $b\to 0$. 
This effect was presented as a manifestation of an extra recoil energy induced by the superkick.
In contrast, we show that the cross section decreases as $b \to 0$.
This dependence is mild, and although it is driven by the proximity to the phase vortex,
it cannot be directly attributed to the superkick phenomenon. 

We believe that the origin of the discrepancy lies in the treatment in \cite{Afanasev:2020nur,Afanasev:2021fda} 
of the target hadrons as pointlike particles of negligible localization size.
As we already demonstrated in our previous paper \cite{Ivanov:2022sco} and reconfirmed in Section~\ref{subsection-superkick} 
of the present work, one cannot resort to this approximation when $b$ is too small.
Disregarding $\sigma_{2\perp}$ is not a way out:
if one keeps $b \ll \sigma_{1\perp}$ and decreases $\sigma_{2\perp}$,
one will unavoidably enter the regime in which the impulse approximation condition
\eqref{impulse-conditions} no longer hold. This will lead to a rapid spreading 
of the wavepacket over the duration of the collision, which unavoidably weakens the effect. 

\subsection{Other scattering processes}

The above results are obtained for the scalar particle collision $\phi\phi\to \Phi\Phi$
which proceeds via the pointlike quartic interaction and results in a constant invariant amplitude ${\cal M}$.
In realistic processes, the invariant amplitude depends on the initial and final state kinematics,
as well as on the polarization states of fermion and vector fields.
Nevertheless, the above analysis can be applied to most of these cases, 
at least within the paraxial approximation.

First, suppose that the plane-wave amplitude ${\cal M}$ entering the expression for ${\cal I}$ in \eqref{cal-I} 
depends only on the total energy and momentum but not on $k_1$ and $k_2$ individually.
Then, ${\cal M}$ can be taken out of the integral, and the analysis of the cross section can be conducted
along the same lines as before.
In particular, for the cross section far above the threshold we still recover the same expression
\eqref{dsigma-like-PW}. This applies, for example, 
to a heavy resonance production $\phi\phi \to \Phi \to \phi\phi$.
If a resonance occurs in the threshold region, we can still rely on Eq.~\eqref{tilde-beta} but with $|{\cal M}|^2$ contributing
to the non-trivial $M_{\rm inv}$ dependence.

Talking specifically about the threshold behavior in $e^+ e^- \to \mu^+\mu^-$ annihilation
near the muon pair production threshold, one needs to take into account the Sommerfeld enhancement
of the cross section due to the Coulomb attraction of the final pair
as well as the sub-threshold production of bound $\mu^+\mu^-$ states. 
It would be interesting to perform this calculation for the LG vs Gaussian wave packet collision
and track the competition among these effects.

Even if the plane wave amplitude ${\cal M}$ depends on the initial momenta,
this dependence is smooth in most cases. Then, for sufficiently narrow momentum-space wave functions,
one could approximate ${\cal M}$ by a constant and take it out of the integral.
The only exception is when ${\cal M}$ varies sharply within the momentum space domain
of the integral ${\cal I}$. This is, for example, the case of the electron elastic scattering
at angles close to the forward peak. 
Whether novel threshold effects in LG vs. Gaussian scattering appear in this situation
can only be answered by a dedicated analysis. 
The examples studied in \cite{Karlovets:2020odl} could be considered a contribution
to this systematic study once extended to the near-threshold region.

\section{Conclusions}

Preparing the colliding particles as compact wave packets of non-trivial shape and phase structure,
such as vortex states,
modifies the cross section in a characteristic way \cite{Jentschura:2010ap,Ivanov:2011kk,Ivanov:2016oue,Karlovets:2016jrd,Karlovets:2018iww,Karlovets:2020odl}.
Novel opportunities for quantum electrodynamics, nuclear, and hadronic processes
offered by so shaping the initial state particles are just beginning to receive attention and appreciation \cite{Ivanov:2022jzh}.
Many of these effects require theoretical treatment which goes beyond the standard calculation procedures used 
in high-energy physics.

In this work, motivated by the recent predictions of strong threshold effects in vortex photon-hadron 
processes in the superkick regime \cite{Afanasev:2020nur,Afanasev:2021fda} and unsatisfied with our previous 
treatment of the superkick phenomenon \cite{Ivanov:2022sco},
we developed the formalism of collisions involving
compact vortex states in the form of Laguerre-Gaussian wave packets.
We drew inspiration from the recent works \cite{Karlovets:2018iww,Karlovets:2020odl}
and, staying within the paraxial approximation, obtained explicit expressions for the
coordinate and momentum wave functions, discussed the parameter choices and dependences,
tracked the time evolution of wave packet collisions, defined the impulse approximation,
and evaluated the toy model cross section in this approximation.
Qualitative discussions of the key physics effects were confirmed by numerical calculations.
With this approach, we were able to rederive the superkick effect 
\cite{barnett2013superkick,Afanasev:2020nur,Afanasev:2021fda} without resorting 
to the artificial procedure used in our previous work \cite{Ivanov:2022sco}.

Equipped with this formalism, we examined the energy behavior of the cross section 
of two heavy particle production in collisions of two light ones prepared as LG and Gaussian wave packets
at the impact parameter $b$.
Far above the threshold, the cross section approaches the plane-wave expression for any wave packet shape,
which is consistent with the previous results \cite{Karlovets:2020odl}.
Near the threshold, we observed $b$-dependent modifications; 
these effects are substantial even in the paraxial approximation and were not dicussed in \cite{Karlovets:2020odl}.
First, due to non-monochromaticity of localized wave packets, 
the sharp threshold is now blurred. However, this effect is not related to the superkick phenomenon
as it exists for Gaussian-Gaussian scattering and does not require the presence of a phase vortex.
Second, our results for the LG vs. Gaussian cross section in the threshold region 
exhibit a dip at $b \to 0$, not an enhancement, if compared with the Gaussian-Gaussian collision cross section.
This dip is driven by the phase vortex but it quickly disappears above the threshold.
We also argued that these results should apply to other processes for which the plane wave invariant amplitude 
varies slowly with the scattering kinematics.

Thus, our results do not support the predictions of \cite{Afanasev:2020nur,Afanasev:2021fda}
and are, in fact, opposite to theirs.

\section*{Acknowledgments}
This work was supported by the Fundamental Research Funds for the Central Universities, Sun Yat-sen University.


\appendix

\section{Evaluation of ${\cal I}$}\label{appendix-I}

Here, we provide details on the evaluation of the integral ${\cal I}$ for the pointlike interaction. 
In order to compute it, we begin with the definition of ${\cal I}$ in \eqref{cal-I},
eliminate $d^3k_2$ via $\delta^{(3)}(\bk_1+\bk_2-\bP)$,
while the remaining energy delta-function is expressed in the form of time integral in the spirit of Eq.~\eqref{delta4}.
Then, the integral takes the form 
\begin{equation}
{\cal I} = \frac{{\cal M}}{(2\pi)^7}\int dt\, e^{iE_f t}\int \frac{d^3k_1}{4E_1E_2} e^{-iE_1t -iE_2t}\varphi_1(\bk_1)\varphi_2(\bP-\bk_1)\,.
\label{integral-I1}
\end{equation}
We work in the paraxial approximation, so that  
$|\bk_1|, |\bP| \ll |p_{iz}|$ as well as $\delta k_{iz} = k_{iz}-p_{iz} \ll |p_{iz}|$.
For the moment, we do not limit ourselves to the center of motion frame,
which means that $P_z = k_{1z}' + k_{2z}'$ can be large but $\Delta P_z = P_z - (p_{1z} + p_{2z})$
is small, $|\Delta P_z| \ll |p_{iz}|$.
Then, we expand the energies $E_1$ and $E_2$ as in \eqref{E-expansion},
omitting the expression in the brackets due to the impulse approximation:
$E_1 + E_2 \approx \varepsilon_1 + \varepsilon_2 + v_1 \delta k_{1z} + v_2 (\Delta P_z - \delta k_{1z})$.
With all the factors written explicitly, the integral takes the form
\begin{eqnarray}
{\cal I} &=& {\cal M}\frac{(4\pi)^{3/2}}{(2\pi)^7\, \sqrt{4\varepsilon_1\varepsilon_2}} 
\sqrt{\sigma_{1\perp}^2\sigma_{1z} \sigma_{2\perp}^2\sigma_{2z}} \frac{1}{\sqrt{\ell!}} e^{- i \bb_\perp \bP_\perp} 
\int dt\, e^{i(E_f- \varepsilon_1 - \varepsilon_2 - v_2 \Delta P_z)t}\nonumber\\
&&\qquad \times\ 
\int d^2k_{1\perp}\, d\delta k_{1z} \, e^{-i(v_1-v_2)\delta k_{1z} t}\, (\sigma_{1\perp}k_{1\perp})^\ell \, e^{i\ell \phi_k}
e^{i\bb_\perp \bk_{1\perp}}\nonumber\\
&&\qquad \times\ \exp\left\{-\frac{1}{2}\biggl[k_{1\perp}^2\sigma_{1\perp}^2 + (\bP_\perp - \bk_{1\perp})^2\sigma_{2\perp}^2
+ (\delta k_{1z})^2\sigma_{1z}^2 + (\Delta P_z - \delta k_{1z})^2\sigma_{2z}^2\biggr]\right\}\,.\label{I-int-4}
\end{eqnarray}
In this expression, the longitudinal and transverse momentum integrals factorize:
\begin{equation}
{\cal I} =  {\cal M}\frac{(4\pi)^{3/2}}{(2\pi)^7\, \sqrt{4\varepsilon_1\varepsilon_2}} 
\sqrt{\sigma_{1\perp}^2\sigma_{1z} \sigma_{2\perp}^2\sigma_{2z}} \frac{1}{\sqrt{\ell!}} e^{- i \bb_\perp \bP_\perp} 
\cdot {\cal I}_\perp \cdot \int dt\, e^{i(\delta E - v_2 \Delta P_z)t} \cdot {\cal I}_L(t)\,,\label{I-int-5}
\end{equation}
where $\delta E \equiv E_f- \varepsilon_1 - \varepsilon_2$.
The transverse integral is
\begin{eqnarray}
{\cal I}_\perp = 
e^{-P_\perp^2\sigma_{2\perp}^2/2} \int d^2 k_{1\perp}\, (\sigma_{1\perp}k_{1\perp})^\ell \, e^{i\ell \phi_k}
e^{i\bb_\perp \bk_{1\perp} + \bk_{1\perp} \bP_\perp \sigma_{2\perp}^2}\, e^{-k_{1\perp}^2(\sigma_{1\perp}^2+\sigma_{2\perp}^2)/2}\,.
\end{eqnarray}
After some algebra, its square can be written as
\begin{eqnarray}
|{\cal I}_\perp|^2 &=& (2\pi)^2 \frac{1}{(\sigma_{1\perp}^2+\sigma_{2\perp}^2)^2}\left(\frac{\sigma^2_{1\perp}\sigma^2_{2\perp}}{(\sigma_{1\perp}^2+\sigma_{2\perp}^2)^2}\right)^{\ell}
\left[\frac{b_\perp^2}{\sigma_{2\perp}^2} + P_\perp^2 \sigma_{2\perp}^2 + 2b_\perp P_\perp\sin(\varphi_P - \varphi_b)\right]^\ell\nonumber\\[2mm]
&&\qquad \times \exp\left[-\frac{b_\perp^2}{\sigma_{1\perp}^2+\sigma_{2\perp}^2} 
- P_\perp^2\frac{\sigma_{1\perp}^2\sigma_{2\perp}^2}{\sigma_{1\perp}^2+\sigma_{2\perp}^2}\right]\,.
\end{eqnarray}
The longitudinal integral is
\begin{eqnarray}
{\cal I}_L(t) &=& \int d\delta k_{1z}\, e^{-i(v_1-v_2)\delta k_{1z} t}
\exp\left[-\frac{1}{2}(\delta k_{1z})^2\sigma_{1z}^2 
-\frac{1}{2} (\Delta P_z - \delta k_{1z})^2\sigma_{2z}^2\right]\nonumber\\
&=&\sqrt{\frac{2\pi}{\sigma_{1z}^2+\sigma_{2z}^2}} 
	\cdot \exp{\left[ -\frac{(\Delta P_z)^2 \sigma_{1z}^2 \sigma_{2z}^2 +  t^2 (v_1-v_2)^2 + 2 i t(v_1-v_2) \Delta P_z \sigma_{2z}^2}{2(\sigma_{1z}^2 + \sigma_{2z}^2)} \right] }.
\end{eqnarray}
Performing the time integration in \eqref{I-int-5}, we get
\begin{eqnarray}
\int dt\, e^{i(\delta E - v_2 \Delta P_z)t}\, {\cal I}_L(t)
	& = & \frac{2\pi}{|v_1-v_2|} 
	\, \exp{\left[ -\frac{1}{2}(\Delta P)_z^2\frac{\sigma_{1z}^2\sigma_{2z}^2}{\sigma_{1z}^2+\sigma_{2z}^2} 
	-\frac{\sigma_{1z}^2 + \sigma_{2z}^2}{2(v_1-v_2)^2} \left(  \delta E - \Delta P_z \frac{v_1 \sigma_{2z}^2 + v_2 \sigma_{1z}^2}{\sigma_{1z}^2 + \sigma_{2z}^2}  \right)^2  \right] }.
\end{eqnarray}
Combining all the factors, we obtain
\begin{eqnarray} \label{matrix result-appendix}
	|\mathcal{I}|^2 &=& |{\cal M}|^2\frac{2}{(2\pi)^7 \ell!} \frac{\sigma_{1z} \sigma_{2z}}{\varepsilon_1 \varepsilon_2 (v_1-v_2)^2} \cdot 
	\left( \frac{\sigma_{1\perp}^2 \sigma_{2\perp}^2}{(\sigma_{1\perp}^2 + \sigma_{2\perp}^2)^2} \right)^{\ell+1}   
	\left[ \frac{b_\perp^2}{\sigma_{2\perp}^2} + P_\perp^2 \sigma_{2\perp}^2 +2 b_\perp P_\perp \sin{(\varphi_P - \varphi_b)} \right]^\ell 
	\nonumber \\[2mm]
	&& \times 
	\exp\left[  - \frac{b_\perp^2}{\sigma_{1\perp}^2+\sigma_{2\perp}^2} 
	- \frac{ P_\perp^2 \sigma_{1\perp}^2 \sigma_{2\perp}^2}{\sigma_{1\perp}^2+\sigma_{2\perp}^2} 
	- \frac{(\delta E -\Delta P_z v_2)^2 \sigma_{1z}^2 + (\delta E -\Delta P_z v_1)^2 \sigma_{2z}^2}{(v_1-v_2)^2} \right].
\end{eqnarray} 
We have checked that $\int |{\cal I}|^2 d^4P$ coincides with the result \eqref{I2d4P}.
If needed, one can now switch to the average center of motion frame in which $\varepsilon_1 = \varepsilon_2 = E_0$,
$|v_1-v_2|=2v_0$, $\delta E \equiv E_f- \varepsilon_1 - \varepsilon_2$, $\Delta P_z = P_z$.


\begin{thebibliography}{99}%

\bibitem{Peskin:1995ev}
M.~E.~Peskin and D.~V.~Schroeder,
Addison-Wesley, 1995,
ISBN 978-0-201-50397-5

\bibitem{Kotkin:1992bj}
G.~L.~Kotkin, V.~G.~Serbo and A.~Schiller,
Int. J. Mod. Phys. A \textbf{7}, 4707-4745 (1992)
doi:10.1142/S0217751X92002131

\bibitem{Ivanov:2022jzh}
I.~P.~Ivanov,
Prog. Part. Nucl. Phys. \textbf{127}, 103987 (2022)
doi:10.1016/j.ppnp.2022.103987
[arXiv:2205.00412 [hep-ph]].


\bibitem{Karlovets:2016jrd}
D.~Karlovets,
JHEP \textbf{03}, 049 (2017)
doi:10.1007/JHEP03(2017)049
[arXiv:1611.08302 [hep-ph]].

\bibitem{Allen:1992zz}
L.~Allen, M.~W.~Beijersbergen, R.~J.~C.~Spreeuw and J.~P.~Woerdman,
Phys. Rev. A \textbf{45}, 8185-8189 (1992)
doi:10.1103/PhysRevA.45.8185

\bibitem{Bahrdt:2013eoa}
J.~Bahrdt, K.~Holldack, P.~Kuske, R.~M\"uller, M.~Scheer and P.~Schmid,
Phys. Rev. Lett. \textbf{111}, no.3, 034801 (2013)
doi:10.1103/PhysRevLett.111.034801

\bibitem{photonics-review-2017}
C.~Hernández-García, J.~Vieira, J.~T.~Mendonça, L.~Rego, J.~San~Román, L.~Plaja, P.~R.~Ribic, D.~Gauthier, and A.~Picón,
Photonics {\bf 4} (2017), 10.3390/photonics4020028

\bibitem{Nature-Phot-2019}
J.~C.~T.~Lee, S.~J.~Alexander, S.~D.~Kevan, S.~Roy, and B.~J.~McMorran, 
Nature Photonics {\bf 13}, 205 (2019).

\bibitem{Bliokh:2007ec}
K.~Y.~Bliokh, Y.~P.~Bliokh, S.~Savel'ev and F.~Nori,
Phys. Rev. Lett. \textbf{99}, 190404 (2007)
[arXiv:0706.2486 [quant-ph]].

\bibitem{Uchida:2010}
M.~Uchida and A.~Tonomura, Nature {\bf 464}, 737 (2010).

\bibitem{Verbeeck:2010}
J.~Verbeeck, H.~Tian, and P.~Schlattschneider, Nature {\bf 467}, 301 (2010).

\bibitem{McMorran:2011}
B.~J.~McMorran {\it et al.}, Science {\bf 331}, 192 (2011).

\bibitem{clark2015controlling}
Ch.~W.~Clark, R.~Barankov, M.~G.~Huber, M.~Arif, D.~G.~Cory, and D.~A.~Pushin,
Nature {\bf 525}, 504 (2015).

\bibitem{sarenac2019generation}
D.~Sarenac, C.~Kapahi, W.~Chen, Ch.~W.~Clark, D.~G.~Cory, M.~G.~Huber, I.~Taminiau, K.~Zhernenkov and D.~A.~Pushin,
PNAS {\bf 116} (41) 20328 (2019).

\bibitem{Sarenac:2022}
D.~Sarenac, M.~E.~Henderson, H.~Ekinci, C.~W.~Clark, D.~G.~Cory, L.~Debeer-Schmitt, M.~G.~Huber, C.~Kapahi, and
D.~A.~Pushin, Sci. Adv. {\bf 8}, eadd2002 (2022), arXiv:2205.06263 [physics.app-ph].

\bibitem{luski2021vortex}
A.~Luski, Y.~Segev, R.~David, O.~Bitton, H.~Nadler, A.~R.~Barnea, A.~Gorlach, O.~Cheshnovsky, I.~Kaminer, E.~Narevicius,
Science {\bf 373}~(6559) 1105 (2021)
doi:10.1126/science.abj2451

\bibitem{Paggett:2017}
M.~J. Padgett,
Opt. Express \textbf{25}, 11265 (2017).

\bibitem{Knyazev-Serbo:2018}
B.~A. Knyazev, V.~G. Serbo, 
Phys. Uspekhi \textbf{61}, 449 (2018).

\bibitem{Bliokh:2017uvr}
K.~Y.~Bliokh, I.~P.~Ivanov, G.~Guzzinati, L.~Clark, R.~Van Boxem, A.~B\'ech\'e, R.~Juchtmans, M.~A.~Alonso, P.~Schattschneider and F.~Nori, \textit{et al.}
Phys. Rept. \textbf{690}, 1-70 (2017)
doi:10.1016/j.physrep.2017.05.006
[arXiv:1703.06879 [quant-ph]].

\bibitem{Lloyd:2017}
S.~M.~Lloyd, M.~Babiker, G.~Thirunavukkarasu and J.~Yuan,
Rev. Mod. Phys. {\bf 89}, 035004 (2017).

\bibitem{larocque2018twisted}
H.~Larocque, I.~Kaminer, V.~Grillo, G.~Leuchs, M.~J.~Padgett, R.~W.~Boyd, M.~Segev, and E.~Karimi, Contemporary Physics {\bf 59}, 126 (2018).

\bibitem{sarenac2018methods}
D.~Sarenac, J.~Nsofini, I.~Hincks, M.~Arif, Ch.~W.~Clark, D.~G.~Cory, M.~G.~Huber and D.~A.~Pushin,
New J. Phys. {\bf 20} 103012 (2018).

\bibitem{Jentschura:2010ap}
U.~D.~Jentschura and V.~G.~Serbo,
Phys. Rev. Lett. \textbf{106}, no.1, 013001 (2011)
doi:10.1103/PhysRevLett.106.013001
[arXiv:1008.4788 [physics.acc-ph]].

\bibitem{Jentschura:2011ih}
U.~D.~Jentschura and V.~G.~Serbo,
Eur. Phys. J. C \textbf{71}, 1571 (2011)
doi:10.1140/epjc/s10052-011-1571-z
[arXiv:1101.1206 [physics.acc-ph]].

\bibitem{Ivanov:2011kk}
I.~P.~Ivanov,
Phys. Rev. D \textbf{83}, 093001 (2011)
doi:10.1103/PhysRevD.83.093001
[arXiv:1101.5575 [hep-ph]].

\bibitem{Karlovets:2012eu}
D.~V.~Karlovets,
Phys. Rev. A \textbf{86}, 062102 (2012)
doi:10.1103/PhysRevA.86.062102
[arXiv:1206.6622 [hep-ph]].

\bibitem{Afanasev:2019rlo}
A.~V.~Afanasev, D.~V.~Karlovets and V.~G.~Serbo,
Phys. Rev. C \textbf{100}, no.5, 051601 (2019)
doi:10.1103/PhysRevC.100.051601
[arXiv:1903.12245 [nucl-th]].

\bibitem{Afanasev:2021uth}
A.~V.~Afanasev, D.~V.~Karlovets and V.~G.~Serbo,
Phys. Rev. C \textbf{103}, no.5, 054612 (2021)
doi:10.1103/PhysRevC.103.054612
[arXiv:2102.10380 [nucl-th]].


\bibitem{Ivanov:2012na}
I.~P.~Ivanov,
Phys. Rev. D \textbf{85}, 076001 (2012)
doi:10.1103/PhysRevD.85.076001
[arXiv:1201.5040 [hep-ph]].

\bibitem{Ivanov:2016oue}
I.~P.~Ivanov, D.~Seipt, A.~Surzhykov and S.~Fritzsche,
Phys. Rev. D \textbf{94}, no.7, 076001 (2016)
doi:10.1103/PhysRevD.94.076001
[arXiv:1608.06551 [hep-ph]].

\bibitem{Karlovets:2016dva}
D.~V.~Karlovets,
EPL \textbf{116}, no.3, 31001 (2016)
doi:10.1209/0295-5075/116/31001
[arXiv:1608.08858 [hep-ph]].

\bibitem{Ivanov:2019vxe}
I.~P.~Ivanov, N.~Korchagin, A.~Pimikov and P.~Zhang,
Phys. Rev. Lett. \textbf{124}, no.19, 192001 (2020) 
doi:10.1103/PhysRevLett.124.192001
[arXiv:1911.08423 [hep-ph]].

\bibitem{Ivanov:2020kcy}
I.~P.~Ivanov, N.~Korchagin, A.~Pimikov and P.~Zhang,
Phys. Rev. D \textbf{101}, no.9, 096010 (2020) 
doi:10.1103/PhysRevD.101.096010
[arXiv:2002.01703 [hep-ph]].

\bibitem{Afanasev:2020nur}
A.~Afanasev, C.~E.~Carlson and A.~Mukherjee,
Phys. Rev. Res. \textbf{3}, no.2, 023097 (2021)
doi:10.1103/PhysRevResearch.3.023097
[arXiv:2007.05816 [quant-ph]].

\bibitem{Afanasev:2021fda}
A.~Afanasev and C.~E.~Carlson,
Annalen Phys. \textbf{534}, no.3, 2100228 (2022)
doi:10.1002/andp.202100228
[arXiv:2105.07271 [hep-ph]].

\bibitem{barnett2013superkick}
S.~M. Barnett, M.~Berry,
Journal of Optics \textbf{15}, 12, 125701 (2013).

\bibitem{Afanasev:2017jdf}
A.~Afanasev, V.~G.~Serbo and M.~Solyanik,
J. Phys. G \textbf{45}, no.5, 055102 (2018)
doi:10.1088/1361-6471/aab5c5
[arXiv:1709.05625 [nucl-th]].

\bibitem{berry2013five}
M.~V. Berry, 
Eur. J. Phys. \textbf{34}, 1337 (2013).

\bibitem{Afanasev:2022vgl}
A.~Afanasev, C.~E.~Carlson and A.~Mukherjee,
Phys. Rev. A \textbf{105}, no.6, L061503 (2022)
doi:10.1103/PhysRevA.105.L061503
[arXiv:2202.09655 [quant-ph]].

\bibitem{Ivanov:2022sco}
I.~P.~Ivanov, B.~Liu and P.~Zhang,
Phys. Rev. A \textbf{105}, no.1, 013522 (2022)
doi:10.1103/physreva.105.013522

\bibitem{Karlovets:2018iww}
D.~Karlovets,
Phys. Rev. A \textbf{98}, no.1, 012137 (2018)
doi:10.1103/PhysRevA.98.012137
[arXiv:1803.10166 [quant-ph]].

\bibitem{Karlovets:2020odl}
D.~V.~Karlovets and V.~G.~Serbo,
Phys. Rev. D \textbf{101}, no.7, 076009 (2020)
doi:10.1103/PhysRevD.101.076009
[arXiv:2002.00101 [hep-ph]].

\bibitem{Landau4}
V.~B.~Berestetskii, E.~M.~Lifshitz, and L.~P.~Pitaevskii, 
{\em Quantum Electrodynamics}, Course of Theoretical Physics, Vol. 4
(Pergamon Press, Oxford, 1982).

\bibitem{integrals}
D.~Zwillinger, V.~Moll, I.~S.~Gradshteyn and I.~M.~Ryzhik,
Table of Integrals, Series, and Products (Eighth Edition),
Academic Press (2014).

\bibitem{Maruyama:2017ptl}
T.~Maruyama, T.~Hayakawa and T.~Kajino,
Sci. Rep. \textbf{9}, no.1, 51 (2019)
[arXiv:1710.09369 [hep-ph]].

\end{thebibliography}
\end{document}